\documentclass[preprint2]{aastex}

\begin{document}
	\title{IFU Unit in  SCORPIO-2 Focal Reducer for Integral-Field Spectroscopy on the 6-m Telescope of the SAO RAS}
	
	\author{V.L.~Afanasiev \altaffilmark{1,*}, O.V.~Egorov \altaffilmark{2,1}, A.E.~Perepelitsyn \altaffilmark{1}}
	
	\altaffiltext{1}{Special Astrophysical Observatory (SAO), Russian Academy of Sciences, Nizhnii Arkhyz, 357147 Karachai-Cherkessian Republic, Russia}
	
	\altaffiltext{2}{Sternberg Astronomical Institute, M.~V.~Lomonosov
		Moscow State University, Moscow, 119234 Russia}
	
	\altaffiltext{*}{E-mail: vafan@sao.ru}

\begin{abstract}
We describe the scheme and design features of the new  IFU unit
(Integral Field Unit) meant to perform integral-field spectroscopy
as a part of  SCORPIO-2 focal reducer, which is mounted in the
prime focus of the 6-m telescope of the Special Astrophysical
Observatory of the Russian Academy of Sciences. The design of the
unit  is based on the principle of the formation of array spectra
using a lens raster combined with optical fibers. The unit uses a
rectangular raster consisting of 22$\times$22 square 2-mm diameter
lenses. The image of the object is transferred  by an optical
system with a  $23^\times$ magnification from the focal plane of
the telescope to the plane of the lens raster. The image scale is
$0\farcs75$/lens and the field of view of the instrument has
the size of $16\farcs5\times16\farcs5$. The raster also contains
two extra  $2\times7$ lens arrays to acquire the night-sky spectra
whose images are offset by $\pm$3\arcmin\ from the center. Optical
fibers are used to transform micropupil images into two
pseudoslits located at the IFU collimator entrance. When operating
in the IFU mode a set of volume phase holographic gratings (VPHG)
provides a spectral range of 4600--7300~\AA\ and a resolution
$\lambda/\delta\lambda$ of 1040 to 2800. The quantum efficiency of
SCORPIO-2 field spectroscopy is 6--13\% depending on the grating
employed. We describe the technique of data acquisition and
reduction using IFU unit and report the results of test
observations of the Seyfert galaxy  Mrk\,78 performed on the 6-m
telescope of the Special Astrophysical Observatory of the Russian
Academy of Sciences.
	
\end{abstract}

\keywords{instruments: spectrograph---techniques: data
	analysis---techniques: spectroscopic---techniques: image
	spectroscopy}

\maketitle

\section{INTRODUCTION}
Currently, most of the major telescopes are equipped with
integral-field spectrographs (IFS) designed for field
spectroscopy. The IFS concept was proposed by~\cite{cou82:Afanasiev_n_en} and published by~\cite{van84:Afanasiev_n_en}. Various spectrograph versions
were first made by~{\cite{bac95:Afanasiev_n_en,Arr90:Afanasiev_n_en,afa90:Afanasiev_n_en}}.
In the past century many astronomers viewed field spectroscopy as
a methodologically complicated technique applicable only to a
limited class of tasks limited by the small field of view of such
spectrographs. The new technological advances of recent decades
resulted in the creation of efficient IFS designed for a wide
range of tasks, which we do not list here as this is beyond the
scope of this paper. We only mention dedicated spectrographs used
extensively in the last decades---GMOS~\citep{mur03:Afanasiev_n_en}
and MUSE~\citep{lau06:Afanasiev_n_en} for \mbox{8-m} telescopes and
SAURON~\citep{bac01:Afanasiev_n_en} and
PMAS~\citep{rot05:Afanasiev_n_en} for 3--4-m diameter telescopes.

The first version of IFS used on the 6-m telescope and called MPFS
(Multi Pupil Field Spectrograph) was made in 1989~\citep{afa90:Afanasiev_n_en}.
Various modifications of this spectrograph were used successfully
on the telescope until 2009 and more than 100 papers dedicated to
the study of galaxies of various types and Galactic gaseous
nebulae were published based on the results of observations made
with this instrument. A description of the last version of the
spectrograph can be found in~\cite{afa95:Afanasiev_n_en,afa01:Afanasiev_n_en}. The main
shortcoming of this instrument was its low transmission
(DQE$<$5\%) due to the use of the mirror-lens camera. Currently
the instrument is being redesigned---we plan to mount a lens
camera in it and increase transmission by a factor of 5--6.
\begin{figure*}[t]
 \includegraphics[width=\linewidth]{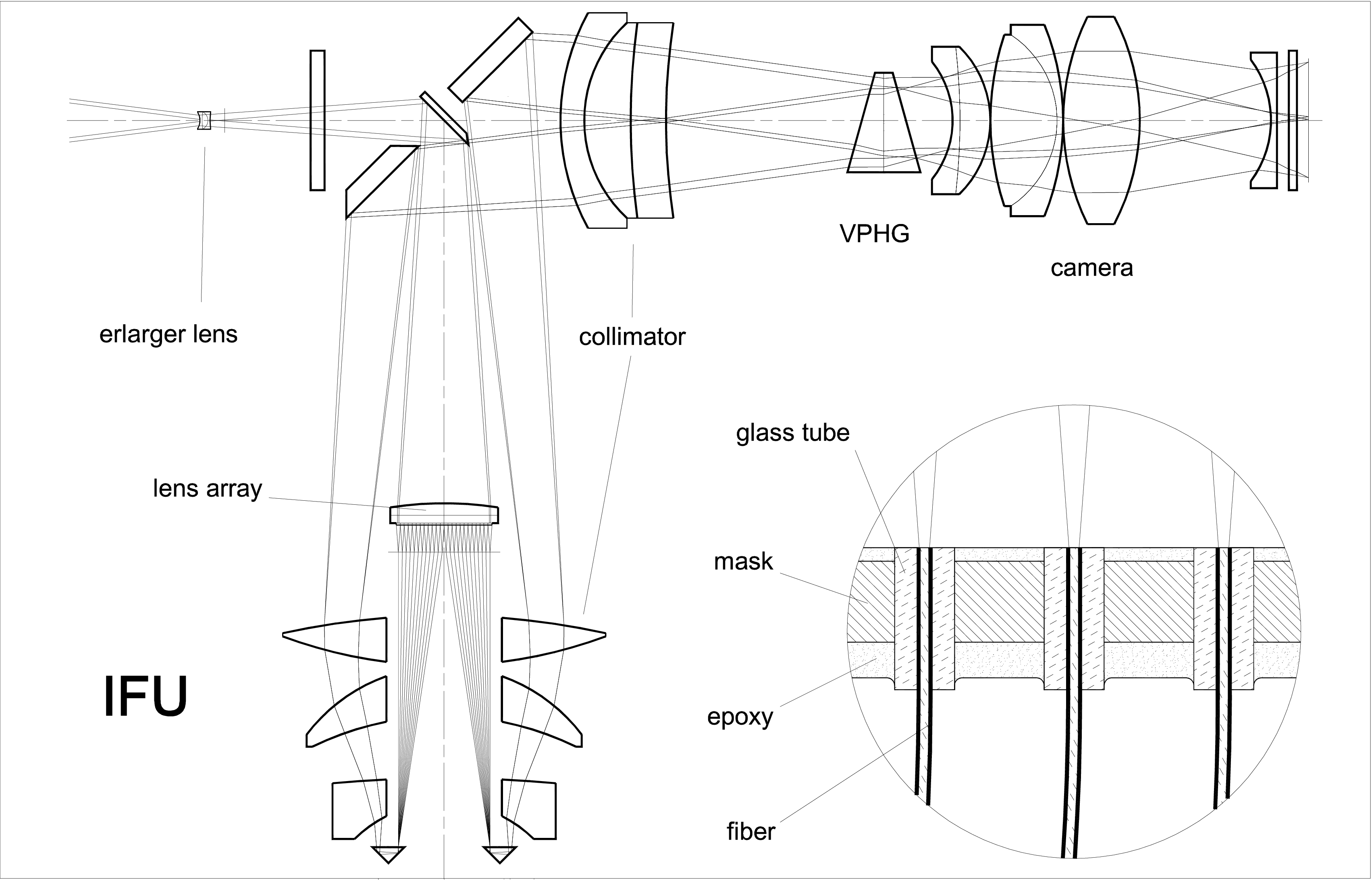}
\caption{Optical scheme of  IFU mode in SCORPIO-2 spectrograph.
The inset image shows the lens raster unit and fibers.}
    \label{optic:Afanasiev_n_en}
\end{figure*}

A new multi-mode SCORPIO spectrograph has been operated at the 6-m
telescope since 2000~\citep{Sco1:Afanasiev_n_en}. The instrument implements
image-acquisition mode, long-slit spectroscopy, multi-object
spectroscopy, spectropolarimetry, and field spectroscopy performed
using a scanning Fabry--Perot interferometer (FPI). Our desire to
expand its capabilities and transition to a larger-format detector
motivated us to begin in 2010 a project aimed at creating the new
multi-mode  SCORPIO-2 spectrograph~\citep{Sco2:Afanasiev_n_en}. A description of
the already implemented capabilities of this spectrograph can be
found on the WEB page of the project\footnote{\url{http://www.sao.ru/hq/lsfvo/devices/scorpio-2/index.html}}.
The new
instrument provides field spectroscopy mode and this paper
describes the IFU mode of SCORPIO-2 spectrograph: design,
technique of data acquisition, and the first results obtained on
the 6-m telescope.

\section{OPTICAL LAYOUT}
\begin{figure*}
\centering
    \includegraphics[width=0.9\linewidth]{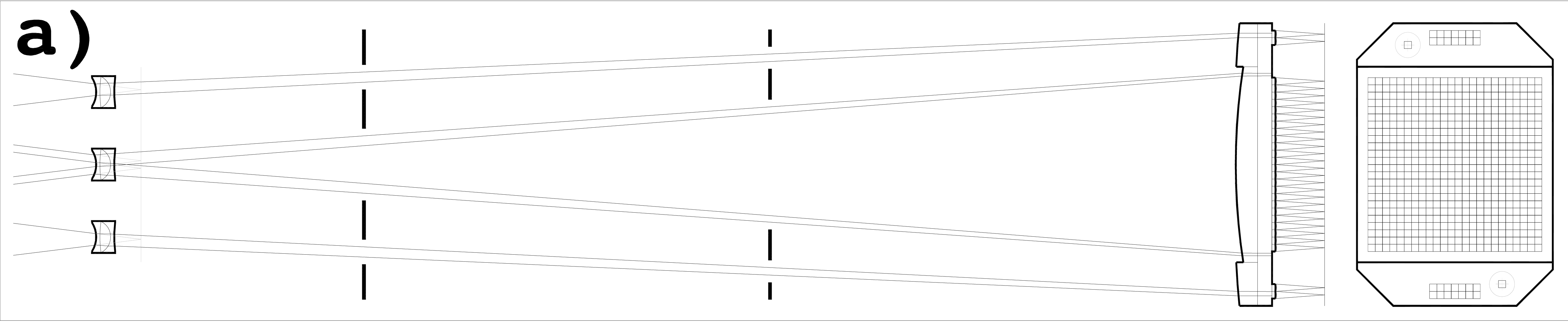}
    \includegraphics[width=0.9\linewidth]{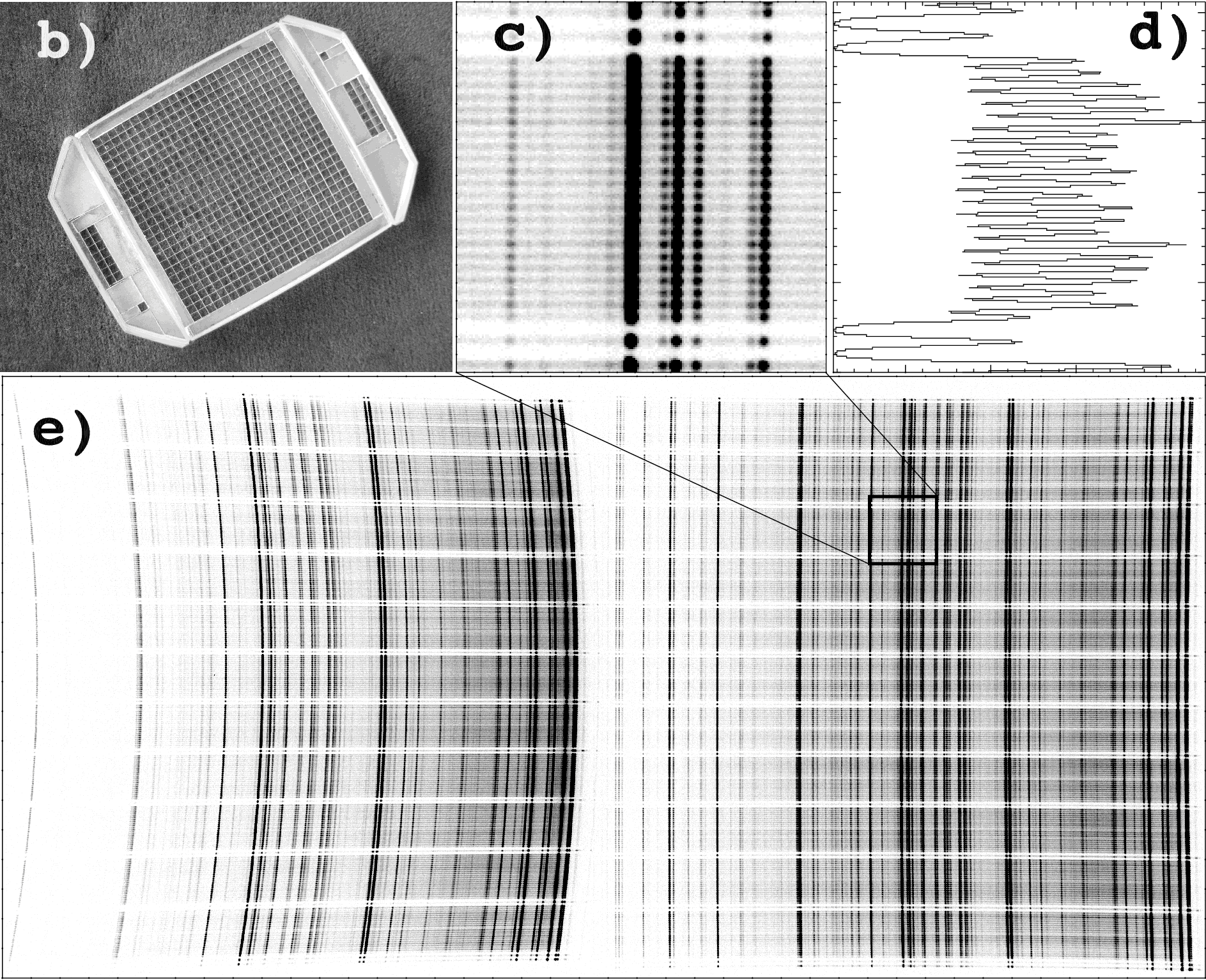}
\caption{Lens raster of  SCORPIO-2  IFU: (a) optical layout of the
image formation and (b) appearance. Examples of spectra acquired
in the IFU mode: (c) selected image fragment, (d) cross section
made across the direction of dispersion, (e) images of the spectra
of the calibrating source (see text for details).}
    \label{IFU-map:Afanasiev_n_en}
\end{figure*}

SCORPIO-2 focal reducer is mounted in the prime focus of the 6-m
telescope of the Special Astrophysical Observatory of the Russian
Academy of Sciences. Unlike most of the field spectrographs of
major telescopes, which are mounted in the secondary foci and have
entrance focal ratios $<$F/8, SCORPIO-2 has an entrance focal ratio
of F/4. This circumstance imposes important constraints on the
design features of the integral-field unit of IFU given the small
size of the spectrograph, which was limited by the volume of the
prime focus cabin. The front segment of SCORPIO-2 (the distance
from the mounting plane of the turret to the slit) is equal to
38~mm, preventing the implementation of technical solutions like
GMOS~\citep{mur03:Afanasiev_n_en} and PMAS~\citep{rot05:Afanasiev_n_en}. We found a solution,
which allows mounting the IFU unit in the spectrograph case while
preserving the possibility of fast switching between operating
modes. Our solution is based on the principle of the use of a lens
raster combined with optical fibers suggested by~\cite{cou82:Afanasiev_n_en}, which we were the first to implement on the 6-m telescope~\citep{afa90:Afanasiev_n_en}. Unlike~\cite{rot05:Afanasiev_n_en}, we
use short fibers with the length of 140--150~mm, which allowed us
to accommodate the IFU unit inside the spectrograph. This solution
reduced the light losses and minimized variations of fiber
transmission due to flexure variations. In the IFU we use another
collimator, which is introduced into the beam instead of the
collimator used for the focal reducer. We show the optical layout
of the spectrograph operating in the IFU mode in Fig.~\ref{optic:Afanasiev_n_en}.
The camera of the spectrograph (6-lens apochromat,
\mbox{$F=109$~mm}, F/2.6, $2\omega=36\degr$) is used in all modes
and produces the image at the entrance of the
\mbox{$4600\times2048$~px}$^2$ EEV42-90 CCD
($1$~px\,=\,$13.5$~$\mu$m). The collimator of the IFU unit (5-lens
acrhomat, $F=240$~mm, F/6, $2\omega=22\degr$) is focal-ratio
matched with the entrance aperture of optical fibers. To reduce
the size, we use a solution where the fiber unit is located inside
the collimator, which has 60$\times$80~mm$^2$ holes in the last
three lenses. Two 56-mm long pseudoslits are projected by the
collimator and camera onto the detector plane. The distance
between the slit centers is equal to 66~mm. The volume phase
holographic grating (VPHG) forms two arrays of spectra in the
detector format. The working portion of the spectrum of each
grating is selected by the interference band filter introduced
into the beam after the magnifier lens.

Fig.~\ref{IFU-map:Afanasiev_n_en} gives the idea of the principle of operation of
the IFU mode of the spectrograph. Panel~(a) shows the optical
layout of the formation of the image on the lens raster. The
central negative lens projects the magnified image of the field
studied from the focal plane of the primary mirror of the 6-m
telescope onto the focal plane of the lens raster. We use a lens
raster consisting of a  $22\times22$ array of $2\times2$~mm$^2$
rectangular lenses with a focal distance of $13.5$~mm. Image
magnification in our case is ~$23^{\times}$,  resulting in the
image scale of $0\farcs75$/lens and a $16\farcs5\times16\farcs5$
field of view. The other two lenses mounted at a distance of
$\pm20$~mm from the center ($\pm3\arcmin$) project images of
night-sky portions onto  $2\times7$ lens rasters. To meet the
telecentricity condition, field lenses are mounted in front of each
lens raster. Fig.~\ref{IFU-map:Afanasiev_n_en}b shows the photo of a lens raster.
Each lens forms a $150-\mu$m diameter image of the telescope
mirror (a mucropupil).

The resulting array of micropupils is transferred onto two
pseudoslits via optical fibers. Fig.~\ref{IFU-map:Afanasiev_n_en}e shows the
spectra of the calibrating source (the continuum and line spectrum
are produced by a quartz and  Ar-Ne-He filled lamp, respectively)
from two pseudoslits. The pseudoslit is reformed so that the
columns of one half of the array of mucripupils would be arranged
sequentially along the slit in groups of 22 micropupils separated
by images of micropupils from night-sky portions.
Figs.~\ref{IFU-map:Afanasiev_n_en}c~and~d demonstrated this arrangement in a
magnified fragment of a portion of IFU spectra.

\subsection{Lens Raster with Optical Fibers}

\begin{figure}[t]
    \includegraphics[width=0.93\linewidth]{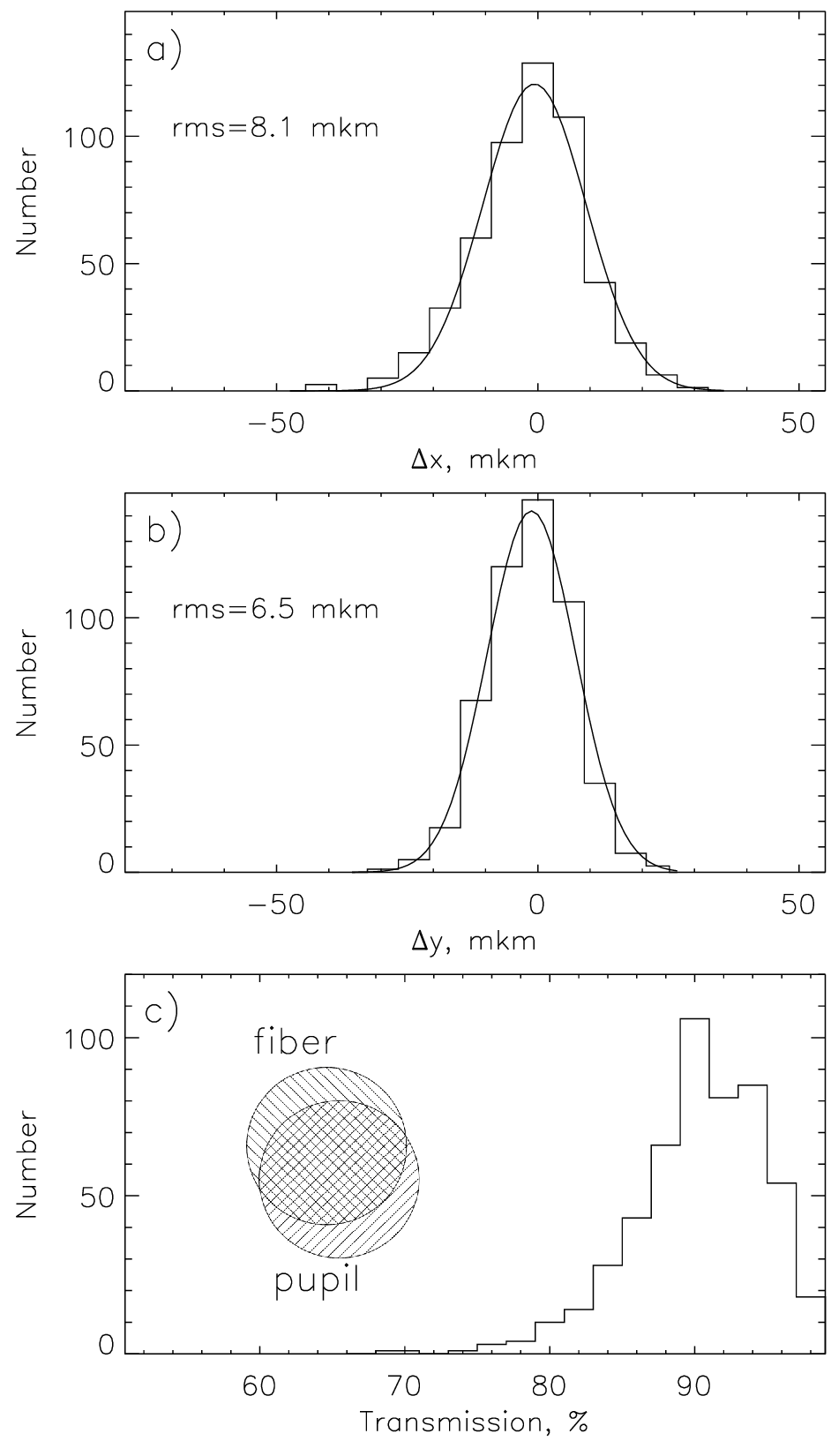}
\caption{Transmission losses of the fiber-lens  IFU unit due to
incorrect mounting of the fibers: (a) and (b)---the distributions
of errors along the  $X$ and $Y$-axes, (c)---Distribution of fiber
transmission, the mean value is  $ 91.5\pm3.3\%$.}
    \label{fibers:Afanasiev_n_en}
\end{figure}

Let us now consider in more detail the design of the unit of lens
raster and fibers. To reform the micropupil array, we use a
custom-made optical fiber that ordered from ``Forc-Photonics
group''\footnote{\url{http://www.forc-photonics.ru/}} with the
following parameters:
\begin{list}{--}{
\setlength\leftmargin{5mm} \setlength\topsep{1mm}
\setlength\parsep{-0.5mm} \setlength\itemsep{2mm} } \item Number
aperture---$0.10\pm0.01$; \item  Core material---quartz glass with
high OH concentration, clear aperture 150~$\mu$m; \item Coating
material---doped quartz glass, diameter 165~$\mu$m; \item
Protective coating---Al, diameter \mbox{195--200~$\mu$m}.
\end{list}
As we pointed out above, the size of the micropupil in our case is
150~$\mu$m, and the maximum aperture of the converging beam after
each lens corresponds to the  F/4 focal ratio and matches the
entrance aperture of the fiber. It is clear that inaccurate
placement of fibers in micropupil positions would result in the
loss of transmission. Precise placement of fibers is a well-known
problem~\citep{rot05:Afanasiev_n_en}, which has several solutions. In our case we
preinstalled the IFU unit on the telescope and obtained a
photographic image of the mucropupil array. We measured the
micropupil positions on the photographic plate and then made a
mask with 805--810-$\mu$m-diameter holes, where we put capillary
glass pipes with the inner and outer diameter of  205--215 and
780--790~$\mu$m, respectively. We put one end of the fiber into
these pipes and the other end, into the 205~$\mu$m-wide gap
between two glass plates (see the inset image in
Fig.~\ref{optic:Afanasiev_n_en}), and fixed them with optical glue. We then
poured epoxy plastic glue onto the fibers with piles, and polished
the fiber ends on both sides. Note that such technology ensures
high quality  of fiber end processing and, as a result, low
Fresnel losses at the entrance and exit of the fibers. After
optical processing and mounting the unit into the case we measured
the fiber positions. The results are shown in Fig.~\ref{fibers:Afanasiev_n_en}.

As is evident from the figure, transmission, which is determined
by the error of fiber placement, is of about 90\%. The actual
transmission also depends on Fresnel losses on optical surfaces
(in our case they are less than 15\%) and chromatic aberrations of
the raster. Computations show that the 150~mm diameter circle
concentrates at least 85\% of the diameter of the polychromatic
micropupil image. We can thus conclude that the transmission of
the lens raster is no less than~65\%. The mechanical structure of
the unit allows adjusting the position of the lens raster relative
to the fiber array using micrometric screws. Adjustment control on
the telescope can also be performed

\subsection{VPHG in the IFU Mode Operation Features}
Volume phased holographic gratings with direct viewing  prisms are
used as dispersing elements in  SCORPIO-2 focal reducer. The VPHG
used in astronomy have high transmission and low scattered
light~\citep{bar00:Afanasiev_n_en}. If $\varphi$ is the  angle at the prism apex
and $\theta$, the angle between the optical axis and the axis of
the collimated beam, then the incidence angle onto the grating,
$\alpha$, can be found from formula $\sin(\varphi-\theta)= n_p
\sin(\varphi-\alpha)$, where $n_p$ is the refractive index of the
prism material. When operating in the common spectroscopy mode
$\theta=0$ (the beam is parallel to the optical axis), while $\theta\neq0$ in the IFU mode. It leads to the difference in the efficiency of VPHG for different slits.

\begin{figure}[t]
    \centering
    \includegraphics[width=\linewidth]{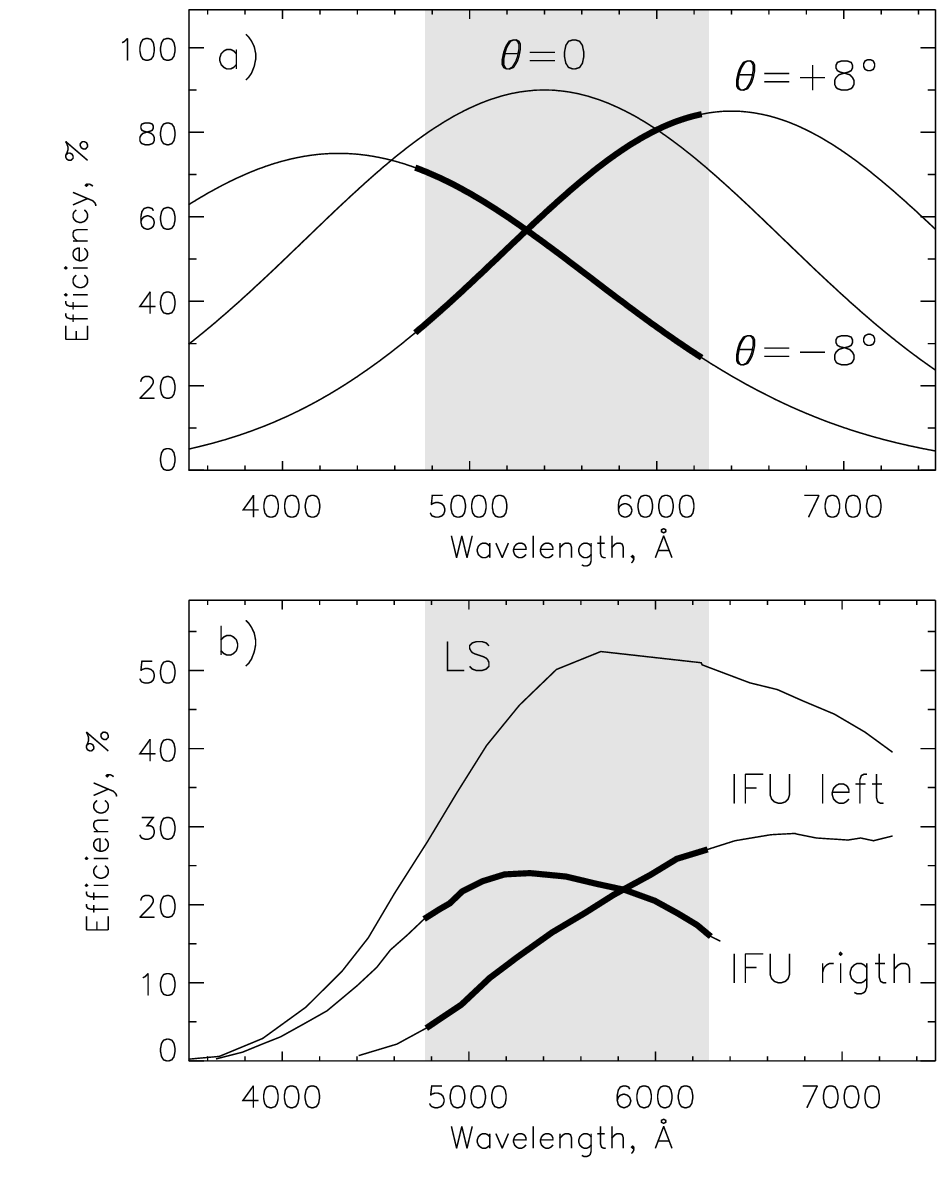}
\caption{Calculated (a) and observed (b) effectiveness of the
VPHG1200@540 grating operating in IFU mode. Gray color is used to
indicate the spectral interval selected by the interference filter
for this grating, \mbox{4900--6150\AA.}}
    \label{eff:Afanasiev_n_en}
\end{figure}

In the Kogelnik approximation~\citep{kog69:Afanasiev_n_en} the efficiency of the
grid in the case of given incidence angle $\alpha$ and wavelength
$\lambda$ is given by the following formula:
\begin{equation}
\eta(\lambda) = \frac{1}{2}\sin^2 \Psi +
        \frac{1}{2}\sin^2(\Psi \cos (\alpha+\beta(\lambda)),
\end{equation}
\begin{displaymath}
where~~~~ \Psi=\frac{\pi \Delta n_g d_g}{\lambda \cos \alpha}.
 \end{displaymath}
Here  $d_g$ is the thickness of the gelatine layer of the grating;
$\Delta n_g$, the amplitude of modulation of the refractive index
$n_g$ of the layer, and $\beta(\lambda)$, the scattering angle for
light with wavelength $\lambda$ onto the grating plane. The
efficiency of the grating reaches maximum at wavelength
$\lambda_{\rm max}$ if $\alpha=\beta(\lambda_{\rm max})$. In this
case angle $\alpha$  is called the Bragg angle and the following
condition is satisfied:
\begin{equation}
\lambda_{\rm max} \nu_g m=n_g \sin(\alpha),
\end{equation}
where $\nu_g$ is the frequency of modulation of the refractive
index of the grating and $m$, the order of scattering, which is
equivalent to the diffraction order of ordinary gratings. Note
that in this case $1/\nu_g$ is equivalent to the number of lines
in ruled gratings; $\alpha$, to the blaze angle, and
$\beta(\lambda)$, to the diffraction angle. Angles between
collimated beams in the IFU and the optical axis (see
Fig.~\ref{optic:Afanasiev_n_en}) are equal to $\pm8\degr$. Fig~\ref{eff:Afanasiev_n_en} shows
the efficiency curves for  VPHG1200@540 grating, which has a
maximum of concentration at $\lambda $ = 5400 \AA\ and $1/\nu_g =
1200$~mm$^{-1}$, computed by formula~(1). We performed our
computations adopting the following values for the grating
parameters: \mbox{$n_g=1.582$}, \mbox{$\Delta n_g=0.0286$}, and
\mbox{$d_g=10.15$~$\mu$m}, and the Bragg angle of $19\fdg2$ in
air. We used a prism made of LK-8 ($n_p=1.473$) glass with angle
$\varphi=36\degr$. As is evident from the figure, theoretical
efficiency curves differ significantly for different slits---on
the average (this corresponds to the middle of the range)
efficiency of the  VPHG when operating in  IFU mode decreases by a
factor of about 1.5 compared to the long-slit (LS) mode. Note that
in the case of ``left''  ($\theta=+8\degr$) and ``right''
($\theta=-8\degr$) slit efficiency increases and decreases with
wavelength, respectively.

To test this conclusion, we compared the spectral fluxes from
quartz lamp in IFU mode with the corresponding fluxes in LS mode
for VPHG1200@540 grating available in SCORPIO-2. We used the data
about the spectrograph efficiency in the LS mode with the account
for spectral sensitivity of the CCD and transmission of the optics
of SCORPIO-2~\citep{Sco2:Afanasiev_n_en}. We compared not only the difference
between the efficiency of VPHG in different modes, but also the
difference between the transmission of optics, which included the
transmission of the linear raster  with light-emitting diodes.
Fig.~\ref{eff:Afanasiev_n_en}b shows the result of a comparison of real spectral
transmission for each slit in the  IFU  and  LS modes. As is
evident from the figure, the efficiency of IFU mode at the center
of the spectral range decreases, on the average, by a factor of
2.5 compared to the LS mode. Given the transmission of the lens
raster with light-emitting diodes, which is on the order
of~\mbox{60--65\%}, this estimate agrees with our computations.

The parameters of the grating used in the IFU mode are listed in
the Table. It gives: grating name (the number in front of the @
symbol indicates the number of lines; the next numbers indicate
the wavelength of maximum concentration in nm, spectral range
selected by the filter in \AA\ px$^{-1}$, and resolution limits
$\lambda/\delta\lambda$ computed by the measured widths of lines.
\begin{table}[t]
\caption{Parameters of gratings in the  IFU mode}
\label{obs-log:Afanasiev_n_en} \label{grating:Afanasiev_n_en}
	\begin{footnotesize}
\begin{tabular}{c|c|c|c}
\hline
Grating  &  Spectral         &Dispersion,&Resolution,\\
         &~coverage, \AA\ & \AA\ px$^{-1}$&$\lambda/\delta\lambda$\\
\hline
VPHG940@600 &4700--7300&0.90--1.15&1044--1269\\
VPHG1200@540&4800--6150&0.81--0.90&1186--1351\\
VPHG1800@510&4600--5400&0.40--0.50&1957--2118\\
VPHG1800@590&5700--6500&0.41--0.51&2375--2453\\
VPHG1800@660&6300--7100&0.42--0.52&2571--2582\\
VPHG2300@520&4930--5630&0.30--0.40&2739--2815\\
\hline
\end{tabular}
\end{footnotesize}
\end{table}

\subsection{Variations of Resolution and Flexure of the Spectrograph when Operated in the IFU Mode}

\begin{figure}[t]
    \centering
    \includegraphics[width=\linewidth]{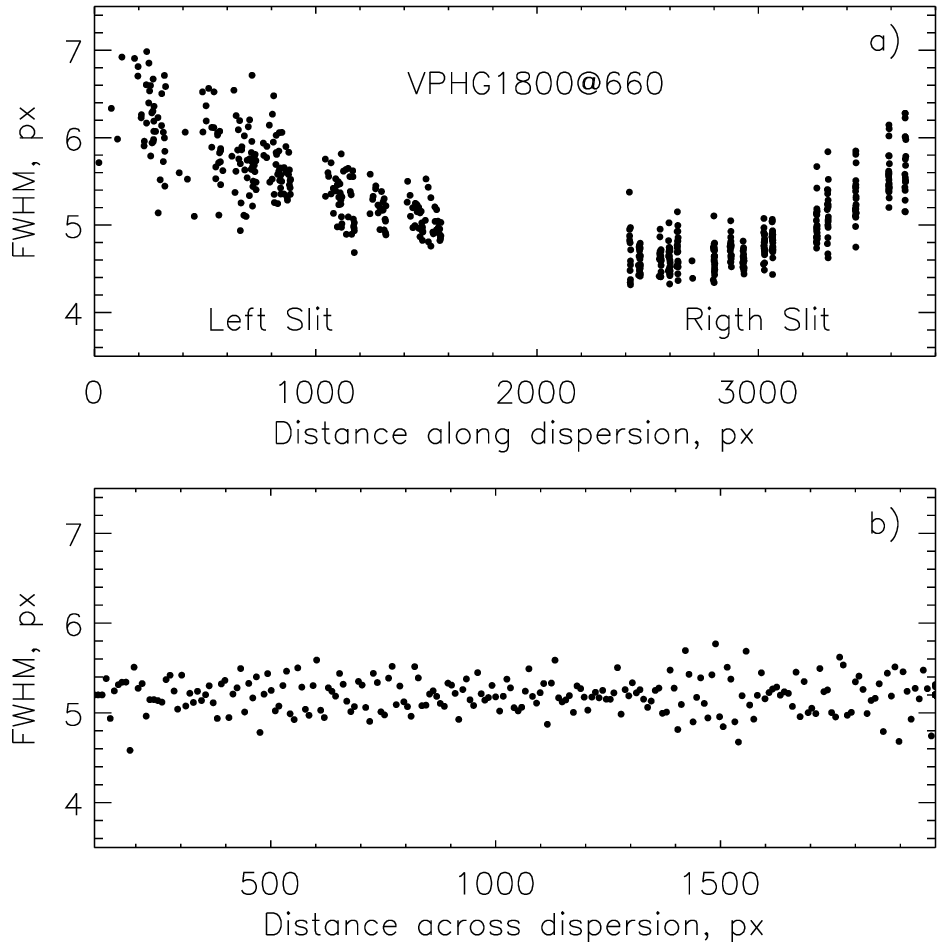}
\caption{Variations of the widths of lines in the comparison
spectrum in the IFU mode for  VPHG1800@660 grating: (a) along and
(b) across the dispersion.}
    \label{fwhm:Afanasiev_n_en}
\end{figure}

The stability of resolution (PSF) across the entire recorded field
in the IFU mode is one of the most important properties of the
instrument. Variations of the line halfwidths in the spectra
formed by individual light-emitting diodes should be small and
vary smoothly across the field of view. This primarily determines
the success of the subtraction of the night-sky spectrum.
Variation of widths across the dispersion affects the quality of
the extraction of spectra and hence spectral reproducibility.
Fig.~\ref{fwhm:Afanasiev_n_en} shows measured FWHM determined from monochromatic
images of individual light-emitting diodes in the spectrum of
neon. The spectrum was acquired with  VPHG1800@660 grating in the
\mbox{6300--7100~\AA} wavelength range, where many bright lines
are recorded. Fig.~\ref{fwhm:Afanasiev_n_en}a shows the variations of the
FWHM of lines along the dispersion. Note that the scatter of
data points is determined by measurement errors. The increase of
FWHM at the edges is determined by the smooth variation of the
angular magnification of the grating and by the decrease of the
camera resolution at the edge of the field of view. The first
circumstance results in appreciable change of the dispersion on
the left slit from 0.52~\AA\ px$^{-1}$ to 0.40~\AA\ px$^{-1}$,
which translates into a change of the line profile in velocity
scale from~110~km\,s$^{-1}$ to~120~km\,s$^{-1}$. Such changes are
acceptable for most of the tasks that require an analysis of the
line profiles of the objects studied. There are no variations of
the profile width along the pseudoslit  (Fig.~\ref{fwhm:Afanasiev_n_en}b) within
the measurement errors. This result is important for choosing
algorithms for extraction of spectra and offers hope for the
realization of good photometric stability of the IFU mode.

\begin{figure}[t]
    \centering
    \includegraphics[width=\linewidth]{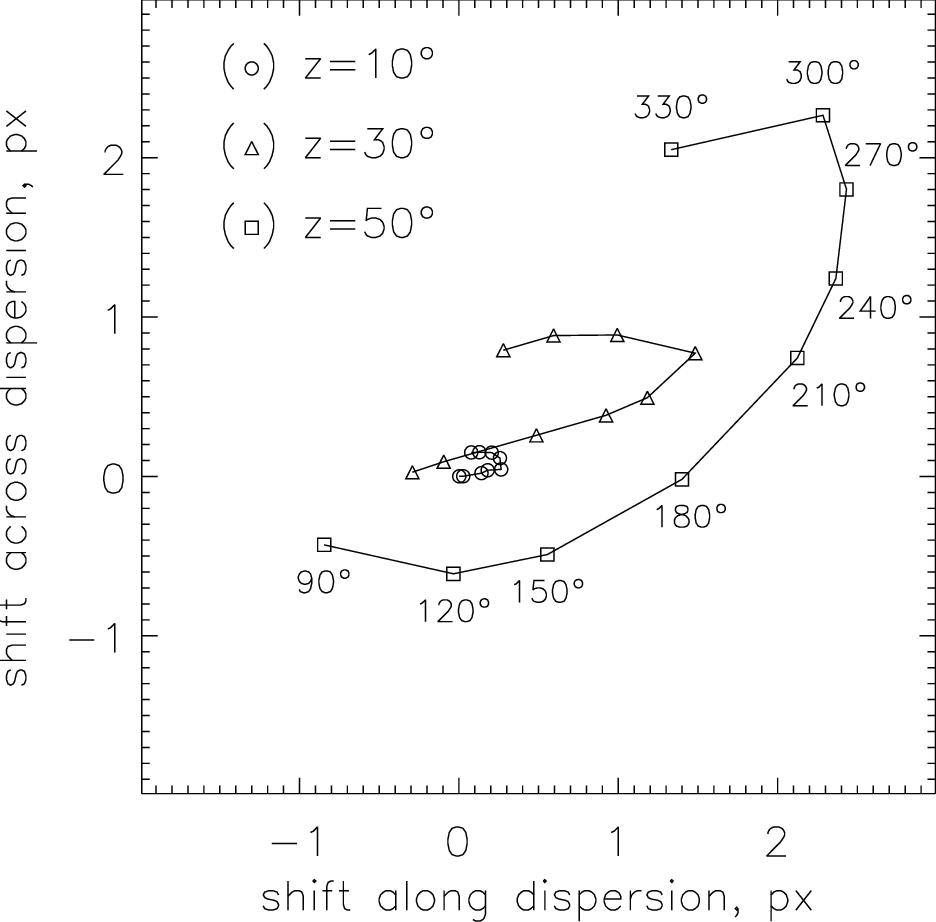}
\caption{Flexures of SCORPIO-2 in the IFU mode at different zenith
angles and in the case of the rotation of the instrument. For
shifts measured at zenith angle 50\degr the corresponding turret
angles are indicated. }
    \label{warp:Afanasiev_n_en}
\end{figure}

Flexures occur in every instrument mounted in the moving focus of
the telescope. Measurement SCORPIO-2 flexures in the LS mode
demonstrate high position stability of the instrument---they do
not exceed 0.8~px (about 10~$\mu$m) in the $z=5\degr$--$60\degr$
interval of zenith angles. A special feature of the  IFU mode is
that only a part of the collimator --- a three-lens assembly and a
diagonal mirror --- is introduced into the beam. The last three
lenses of the collimator, which are located in  the same case with
the fiber-lens unit, are fixed and do not move. In the case of the
change of the zenith angle and rotation of the' instrument this
results in the shift of the line positions. Flexure measurements
in the IFU mode (Fig.~\ref{warp:Afanasiev_n_en}) show that during a  15--20~min
exposure the image shift does not exceed  0.3~pix and does not
degrade the resolution. Given the structure of each image
(intensity ``modulations'' both along the dispersion and along the
pseudoslit), flexure effects are easy to tale  into
account---during reduction of our data each image is reduced to
the unified reference frame to within~0.1~px.

\section{REDUCTION OF OBSERVATIONAL DATA}
\label{reduction:Afanasiev_n_en}

\begin{figure*}
    \centering
    \includegraphics[width=0.49\linewidth]{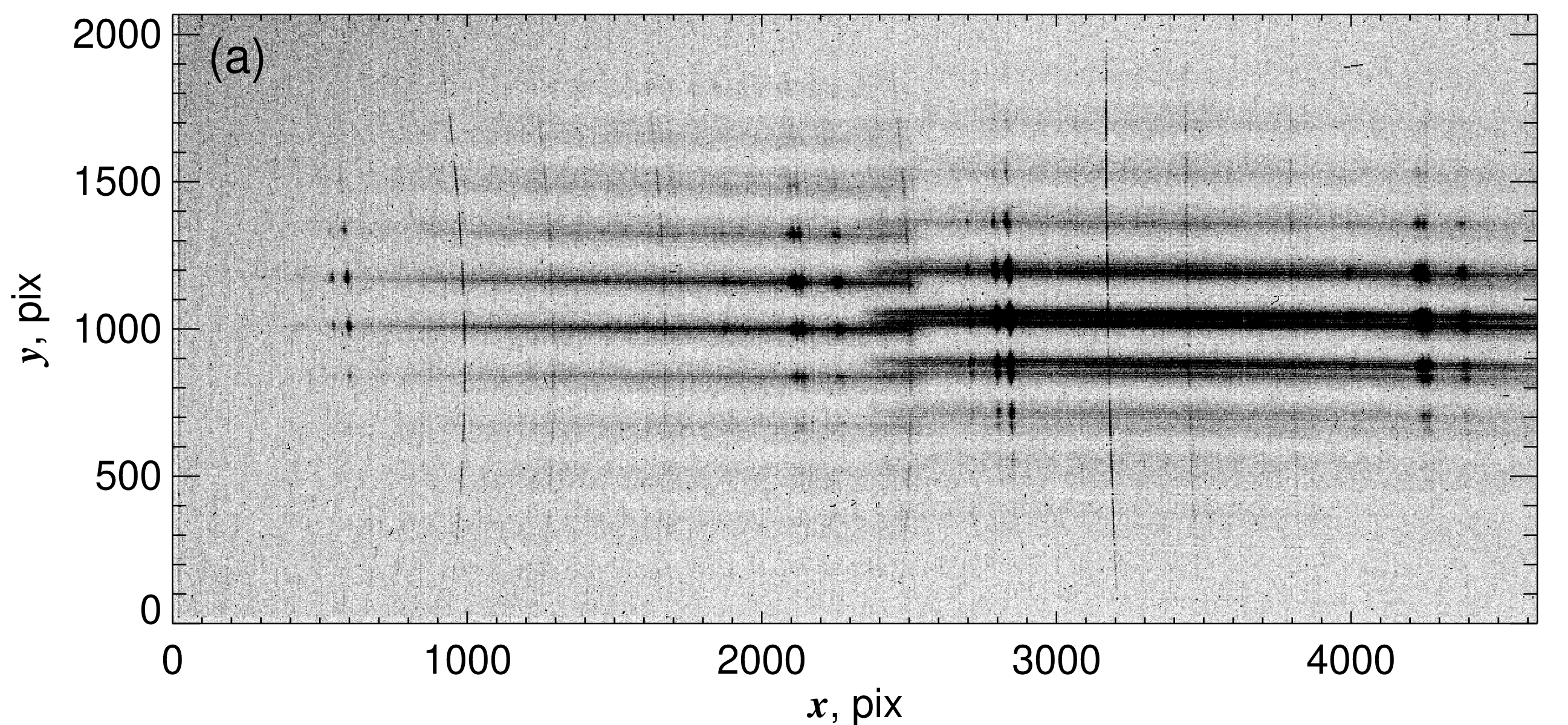}~
    \includegraphics[width=0.49\linewidth]{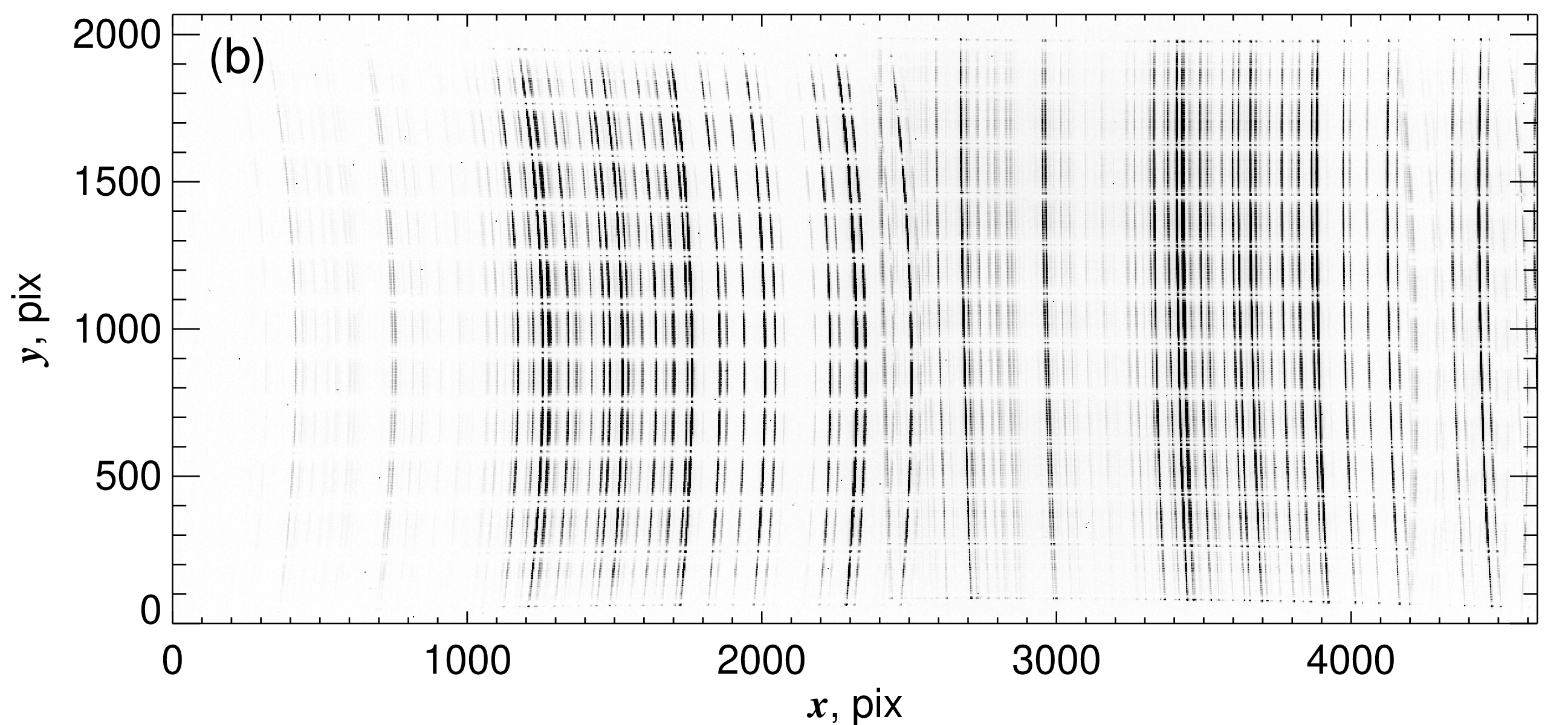}
    \includegraphics[width=0.49\linewidth]{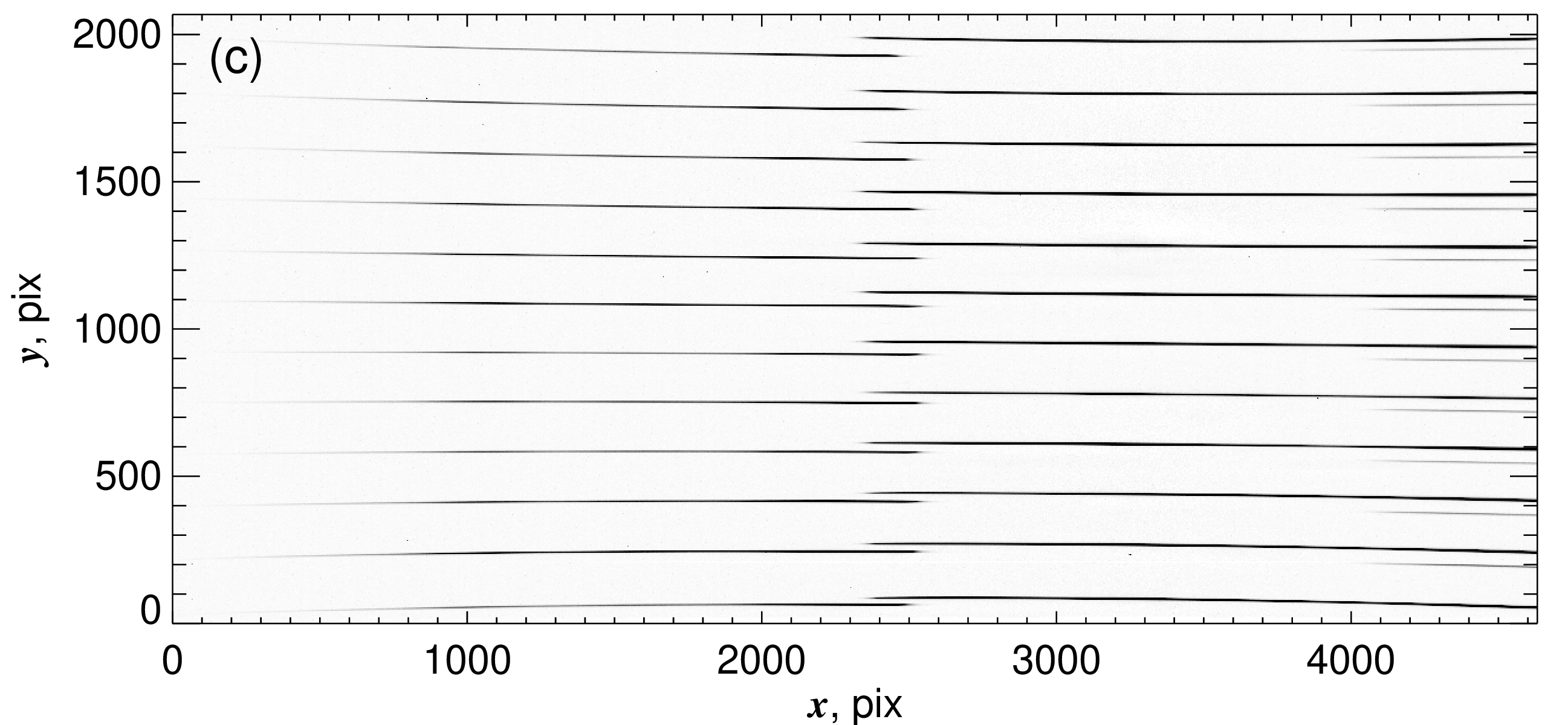}~
    \includegraphics[width=0.49\linewidth]{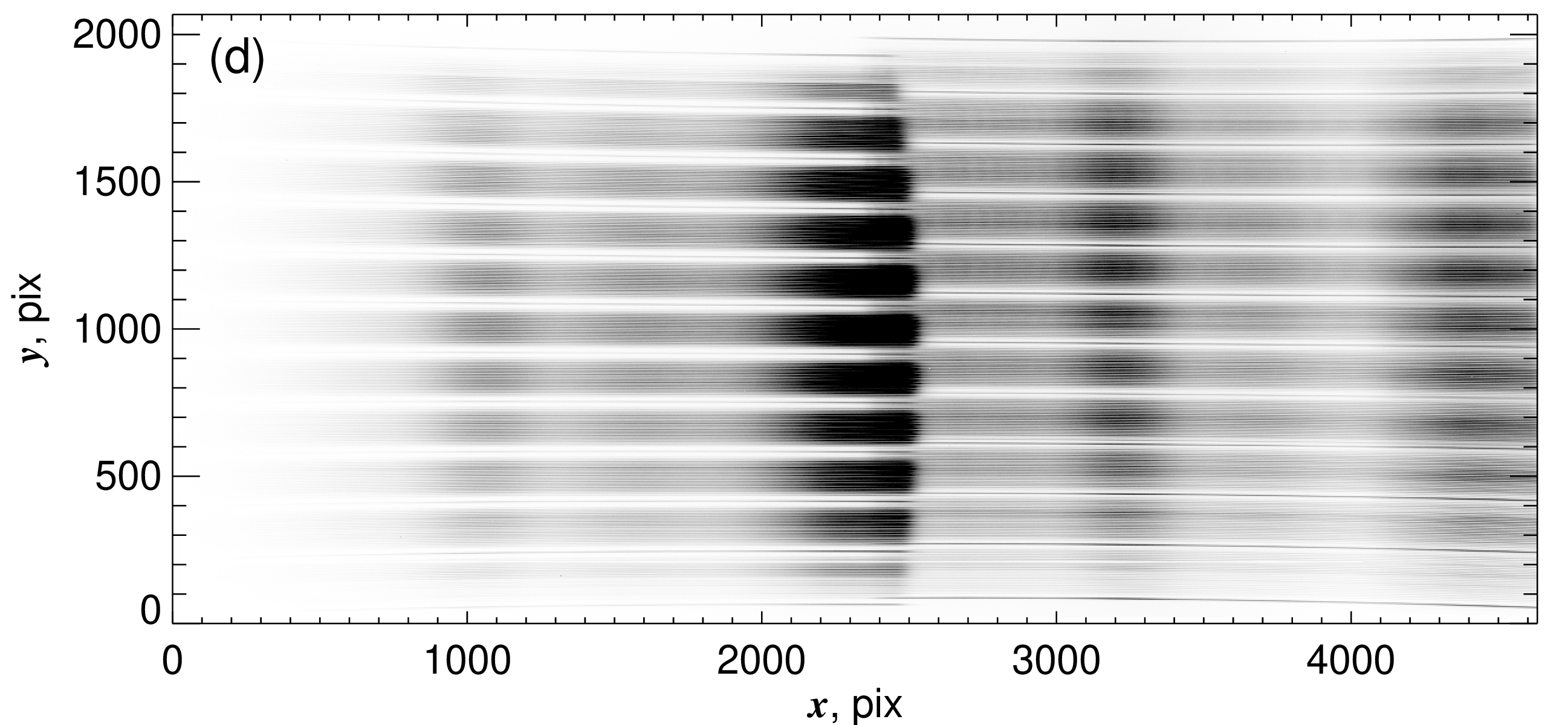}
    \caption{Example of raw frames from the data set acquired with VPHG940@600 grating: (a) \textsc{obj}---the spectrum
    of Mrk\,78; (b) \textsc{neon}---the calibration spectrum of the He-Ne-Ar lamp ; (c)  \textsc{eta}---the standard
    spectrum acquired with  12dots mask; (d) \textsc{flat}---the flatfield spectrum.}
    \label{fig:rawdata:Afanasiev_n_en}
\end{figure*}

\begin{figure}[t]
    \centering
    \includegraphics[width=\columnwidth]{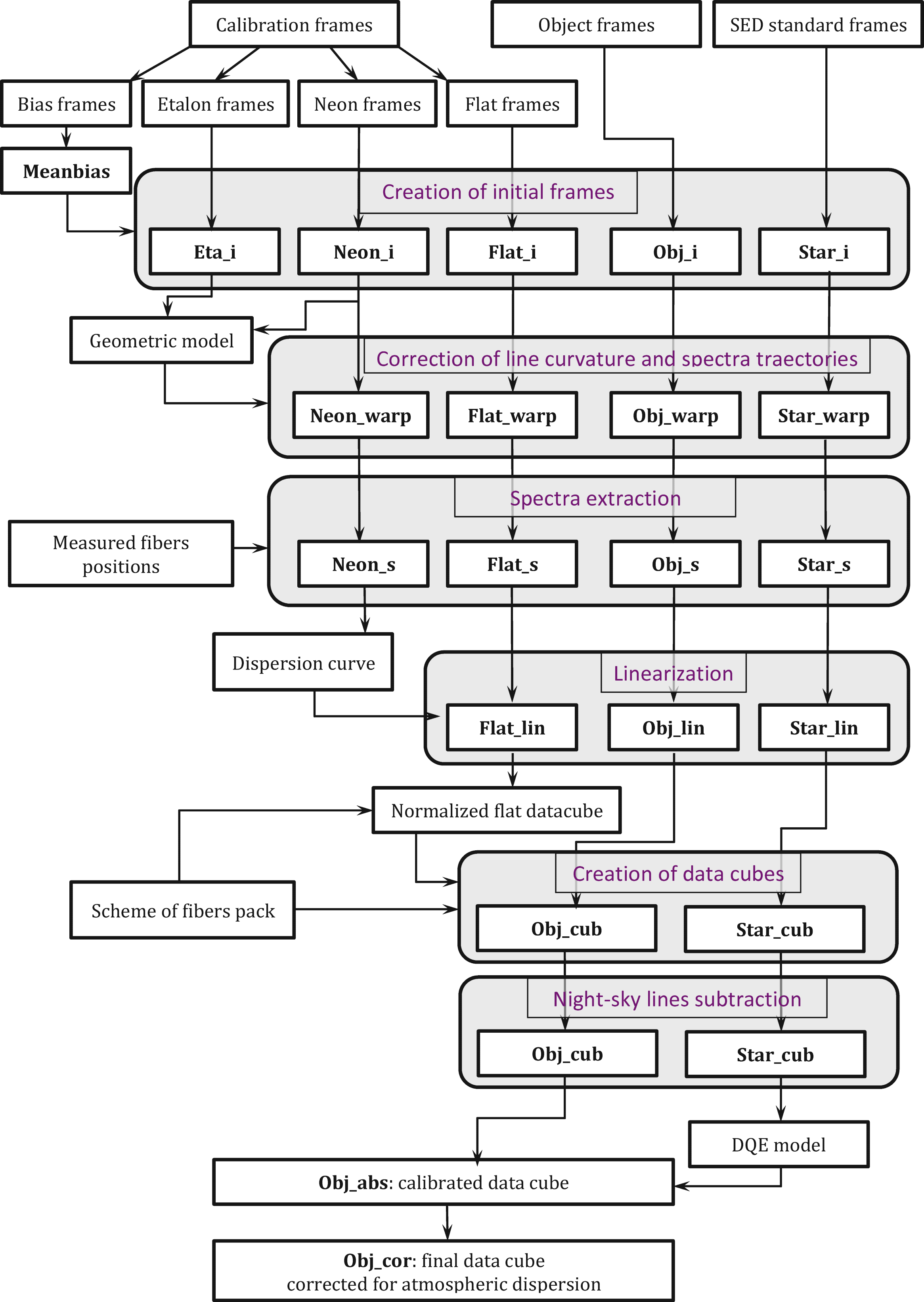}
    \caption{Block diagram of the data reduction  procedure using  IFURED package. The bold font is used to indicate
    the names of the files created at each stage of reduction.}
    \label{fig:scheme:Afanasiev_n_en}
\end{figure}

A complete data set acquired during \mbox{SCORPIO-2} observations with IFU includes: 
\begin{list}{--}{
\setlength\leftmargin{5mm} \setlength\topsep{1mm}
\setlength\parsep{-0.5mm} \setlength\itemsep{2mm} } \item Bias
frames (\textsc{bias}), which are usually taken at the beginning
and at the end of the night; \item Flatfield spectrum
(\textsc{flat}); it is taken immediately before and after the
observations of the object; \item Standard spectrum (\textsc{eta})
obtained by illuminating 12dots mask with flatfield lamp and which
reflects the position of spectra from ``sky'' fibers in the frame;
it is taken once before and once after the observation of the
object; \item Comparison spectrum of the He-Ne-Ar calibrating lamp
(\textsc{neon}); like \textsc{flat}, it is taken before and after
observations of the object; \item Spectra of the observed object
(\textsc{obj}); at least three exposures should be taken to remove
cosmic-ray hits; \item Spectra of spectrophotometric standard
(\textsc{star}) taken shortly before or after observations of the
object at a similar zenith angle; \item Twilight sky spectra
(\textsc{sunsky}); taken in the evening before sunset and in  the
morning before sunrise.
\end{list}

We reduce observational data using standard procedures employed in
3D spectroscopy. Any programs for the analysis of the data from
IFU spectrographs can be used  (e.g., p3D,~\citealt{p3d:Afanasiev_n_en}). We prefer
our own package written in IDL. All reduction stages in IFURED
take into account specific features of the data obtained with  IFU
unit in SCORPIO-2.

Fig.~\ref{fig:rawdata:Afanasiev_n_en} shows raw frames of various types:  (a)
\textsc{obj}, (b) \textsc{neon}, (c) \textsc{eta}, and (d)
\textsc{flat}. During our test observations no \textsc{sunsky} has
been taken and we therefore do not show its example and do not
consider it in our description of the data reduction procedure.

As is evident from Fig.~\ref{fig:rawdata:Afanasiev_n_en}, the spectra from the
left and right slits overlap approximately at the center of the
frame. Furthermore, during observations with VPHG940@600 grating
the second order from the left slit contributes to the red part of
the spectrum obtained from the right slit (the effect is
especially apparent in the \textsc{eta} images,
Fig.~\ref{fig:rawdata:Afanasiev_n_en}c). In this connection when reducing the
data we fix the boundaries of the spectra from the right and left
slits on the acquired frames trimming cutting the overlap areas.

Below in this section we describe the main stages of reduction
using IFURED package\footnote{Given possible important changes in
the algorithm of the program after the publication of this paper
we advice the reader to read its description at the site of the
project: \url{http://www.sao.ru/hq/lsfvo/devices/scorpio-2/index.html}}.
Fig.~\ref{fig:scheme:Afanasiev_n_en} shows the flowchart
diagram of the IFU data reduction procedure, which has the form of
a sequence of actions. As an example let us consider the data
acquired with  VPHG940@600 grating for the Mrk\,78 galaxy.

\subsection{Assembly of Initial 2D Spectra}
\label{sec:red_ini:Afanasiev_n_en} At the initial stage we average all available
\textsc{bias} frames and subtract the resulting \textsc{meanbias}
frame from all individual exposures.

We then apply corrections to take into account possible offsets
(which may be caused by flexures inside the instrument) between
different exposures both within the same set and between different
types of data. We first use cross-correlation to determine the
offset in the direction perpendicular to dispersion between
individual exposures of the object and shift all frames to align
them with the first exposure. We then use a similar procedure to
determine the offset between the flatfield and object spectra and
then align all the remaining frames with the flatfield spectrum.
For some \textsc{obj}, \textsc{neon}, and \textsc{star} exposures
we also determine the offsets along the dispersion and correct
them. In practice offsets usually do not exceed 1 pixel, but
greater offsets may occur between the \textsc{star} frame and the
remaining set of data, because spectrophotometric standard frame
is take at another, albeit similar, position of the telescope.

We stack individual exposures of each data type in pairs. In the
process, cosmic-ray hits are removed. In each frame of the pair
pixels are found where the signal significantly exceeds the number
of counts in the same pixel in the other frame and intensities in
these pixels are replaced by the counts from the other frame of
the pair.

At this stage the spectra from the right and left slits are
separated and the data cube is created with two channels along the
spectral axis. Each channel contains the spectrum from the right
or left slit. The spectra are then reduced separately using
identical procedure.

\subsection{Correction of Geometric Distortions}

The next stage consists in the construction of a geometric model
of he trajectories of spectra from individual fibers in the frame.
The standard spectrum \textsc{eta} and the calibration spectrum of
the \mbox{He-Ne-Ar} lamp \textsc{neon} are analyzed.

\begin{figure}
    \centering
    \includegraphics[width=0.973\linewidth]{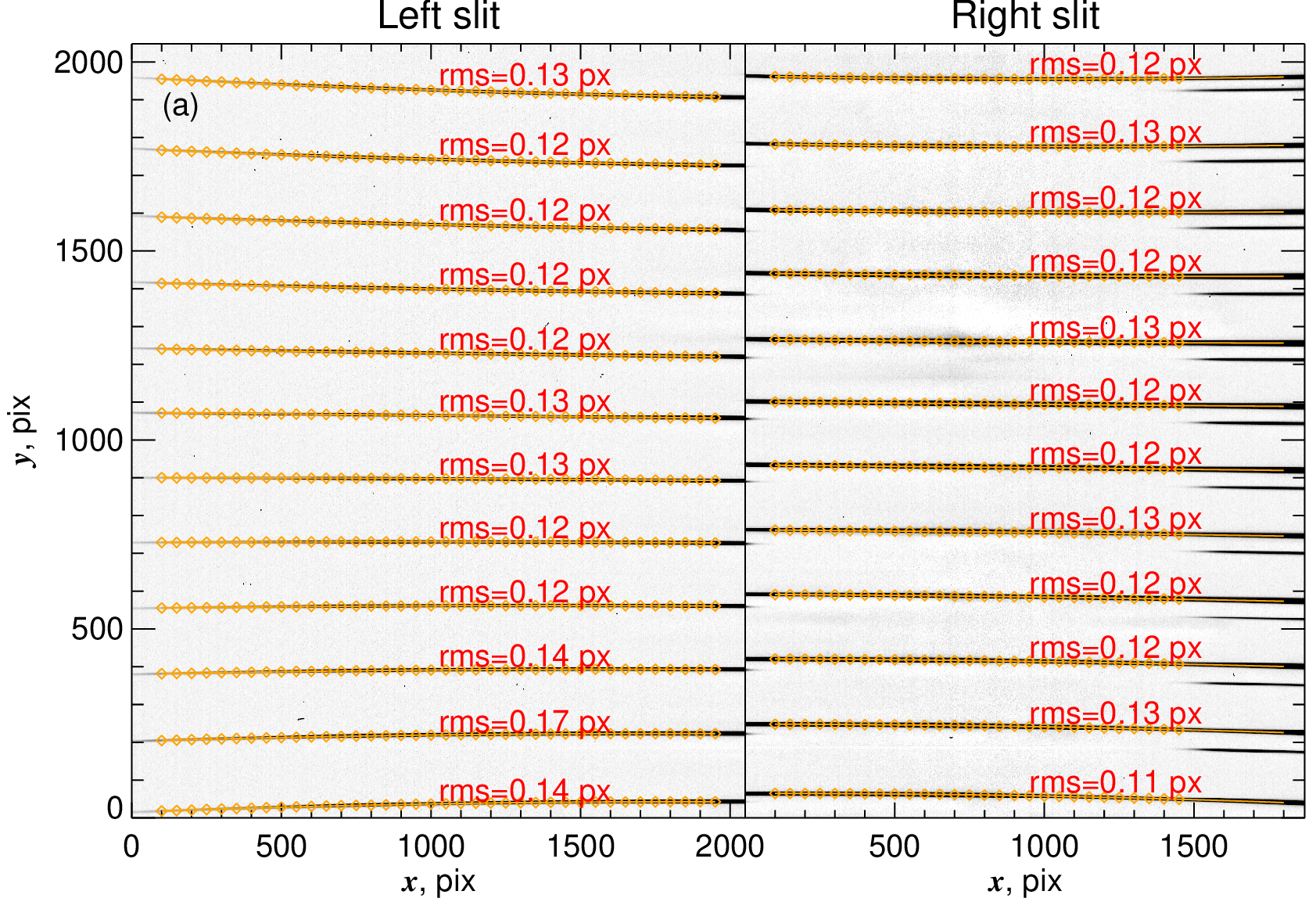}
    \includegraphics[width=0.973\linewidth]{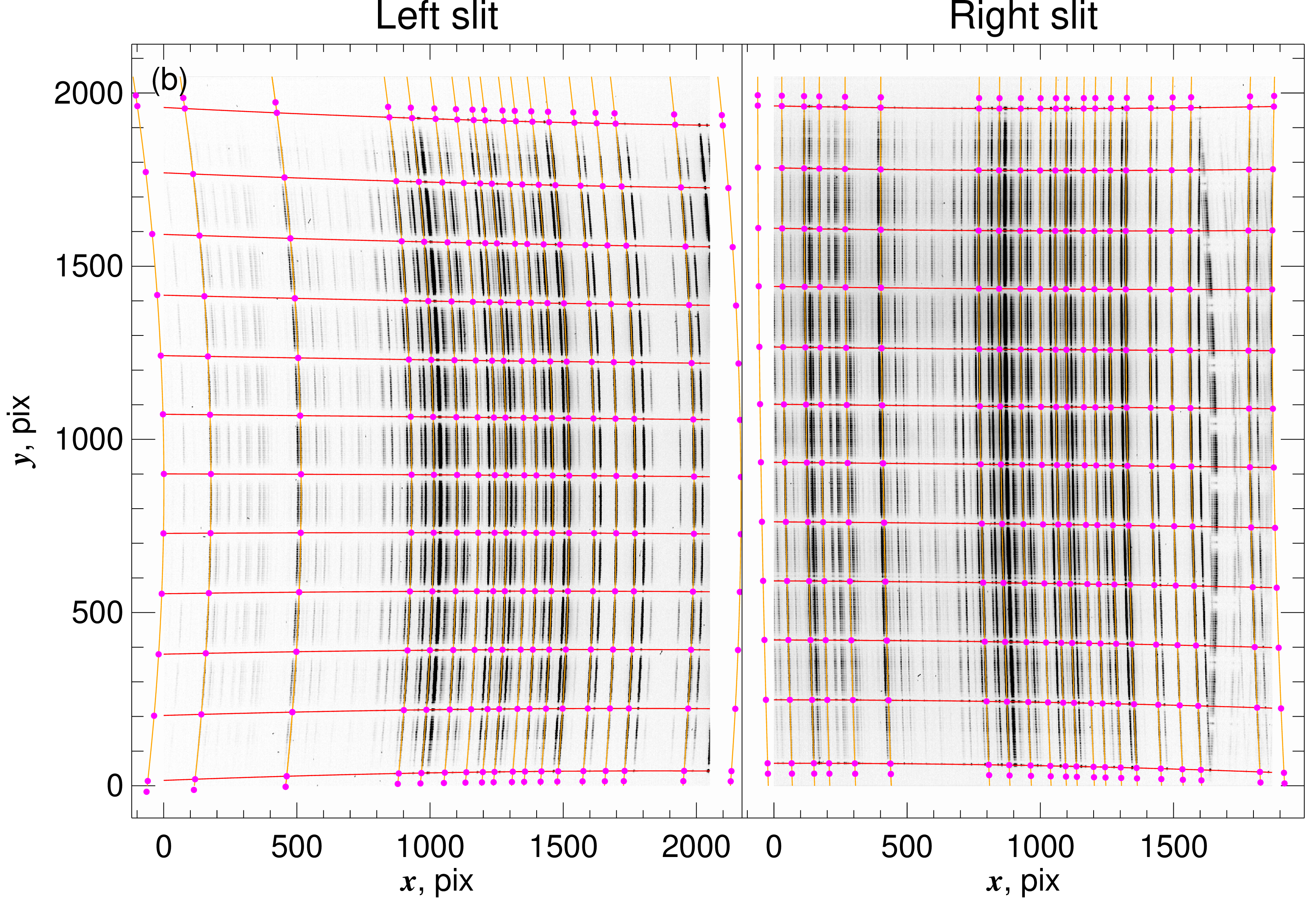}
    \includegraphics[width=0.973\linewidth]{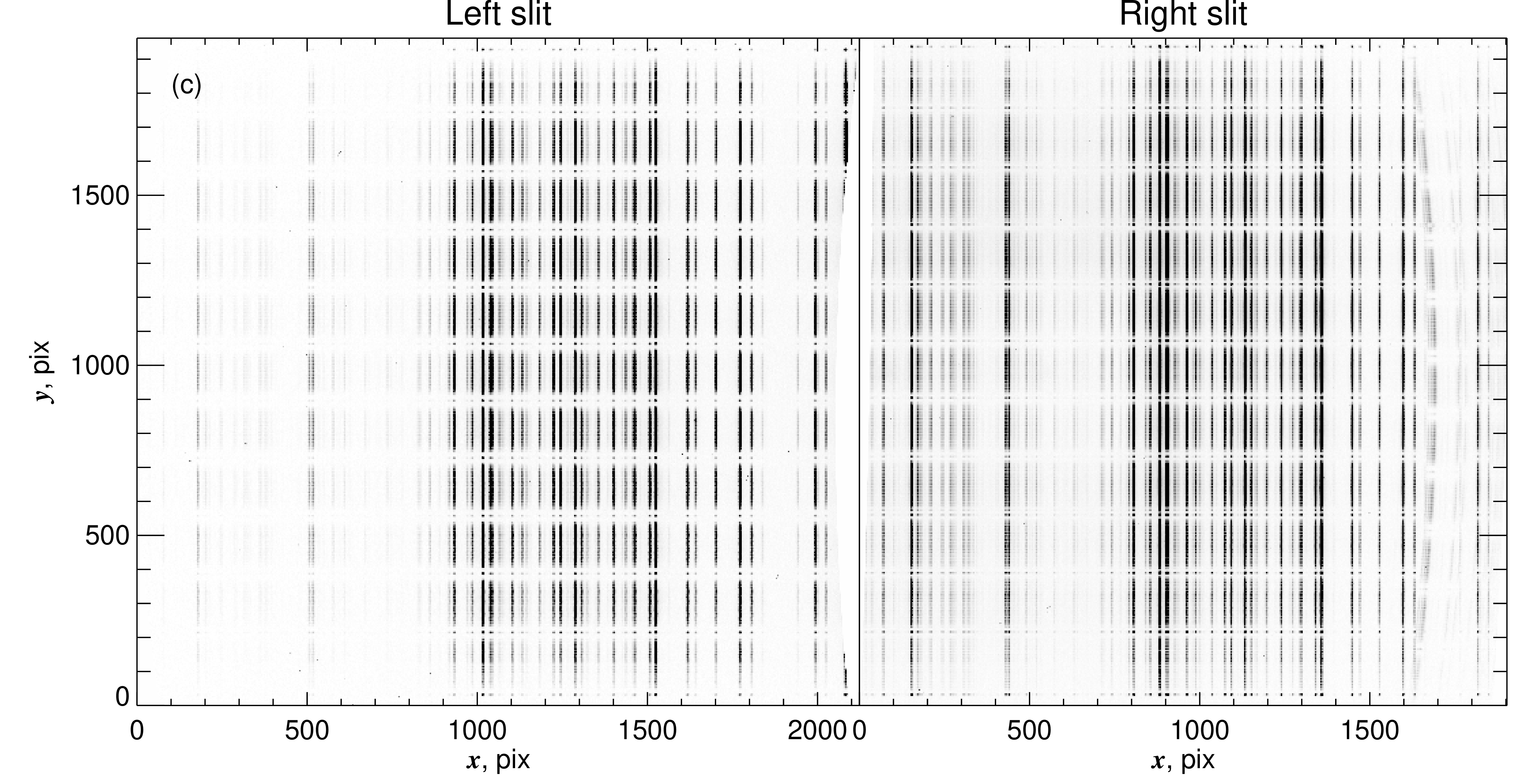}
\caption{Illustration of the procedure of the correction of
geometric distortions. The left and right panels show the results
obtained for the left and right slits, respectively.
(a)---Identification of trajectories of ``sky'' fibers in the
\textsc{eta} spectrum; the accuracy of the procedure for each
fiber is indicated. (b)---The  \textsc{neon} spectrum, identified
lines (orange) and trajectories of  ``sky'' fibers (red), their
intersection points (magenta). Extrapolation is performed beyond
the observed spectrum. (c)---\textsc{neon} after correction for
geometric distortions.}
    \label{fig:geometry:Afanasiev_n_en}
\end{figure}

First, the program searches for trajectories from each  ``sky''
fiber in the \textsc{eta} frame. The images from each slit are
subdivided into several tens of intervals along the dispersion and
the positions of the intensity distribution peaks in the direction
perpendicular to dispersion are searched in each interval. The
portions where the number of peaks is not equal to 12 are
discarded. The coordinates of the remaining points are fitted by a
second-order polynomial. The resulting error of the determination
of the position of fiber trajectories does not  exceed~0.2~pixels
(see Fig.~\ref{fig:geometry:Afanasiev_n_en}a).

\begin{figure*}[t]
	\includegraphics[width=0.495\linewidth]{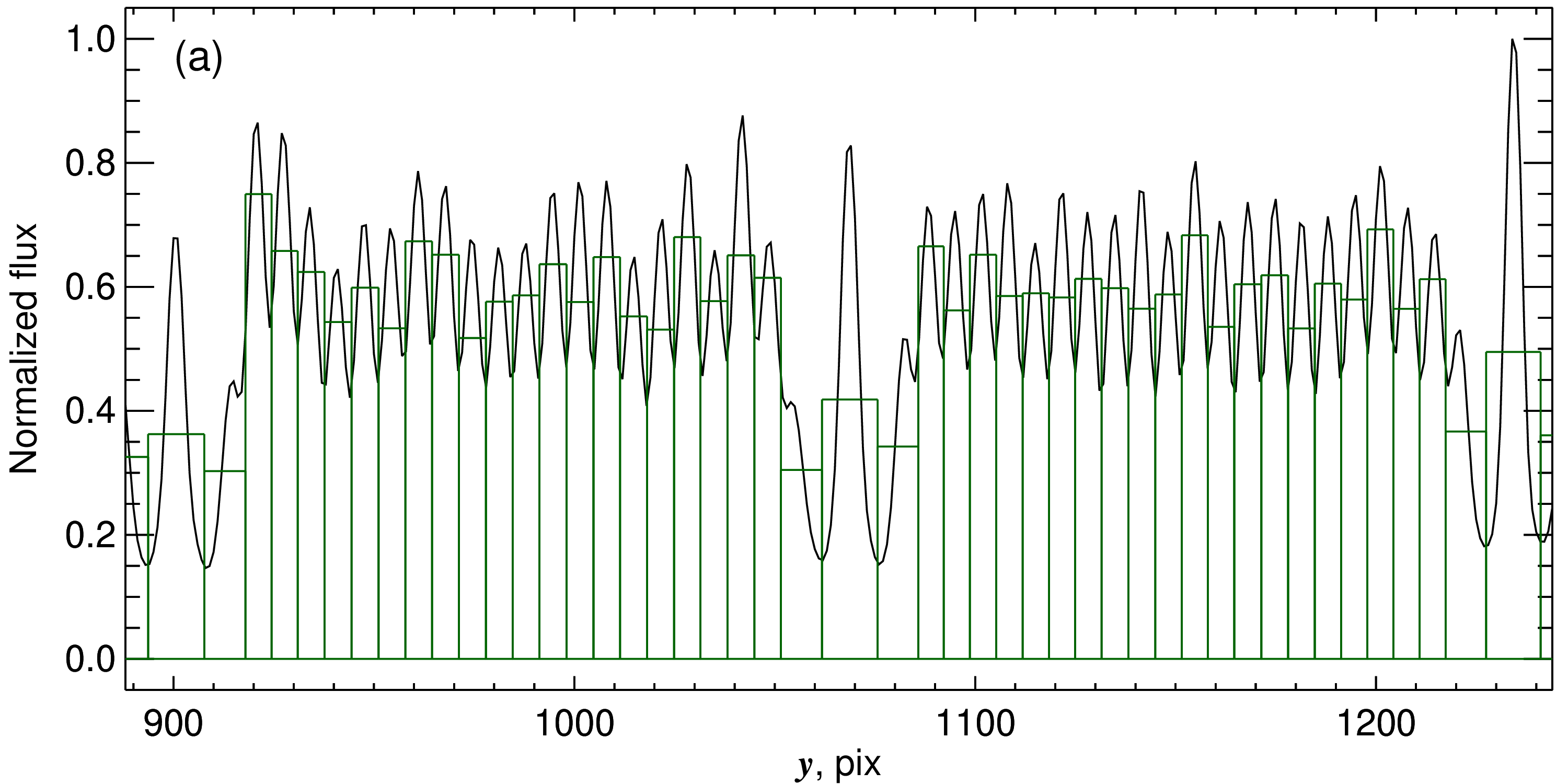}~
	\includegraphics[width=0.495\linewidth]{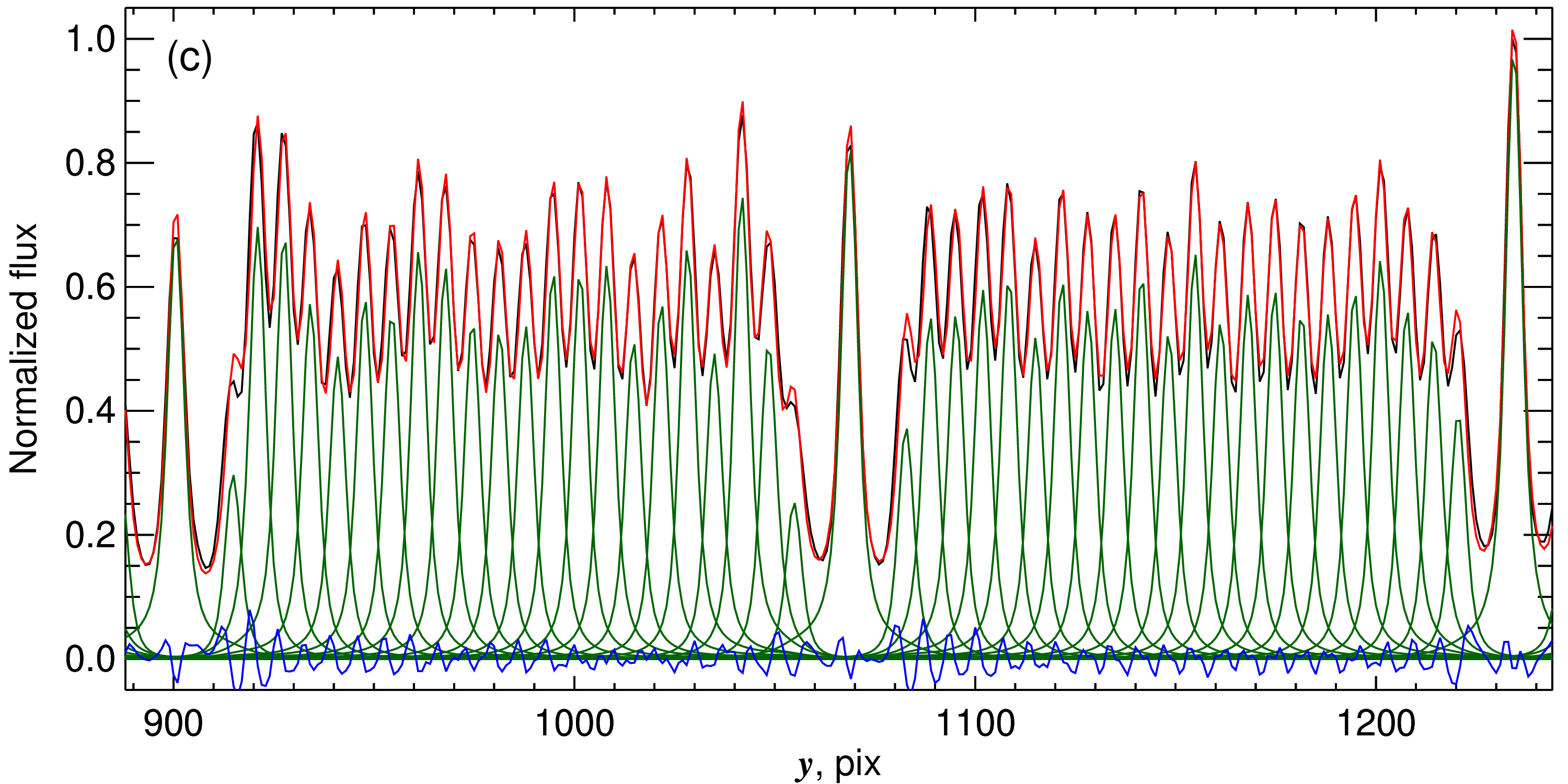}
	\includegraphics[width=0.495\linewidth]{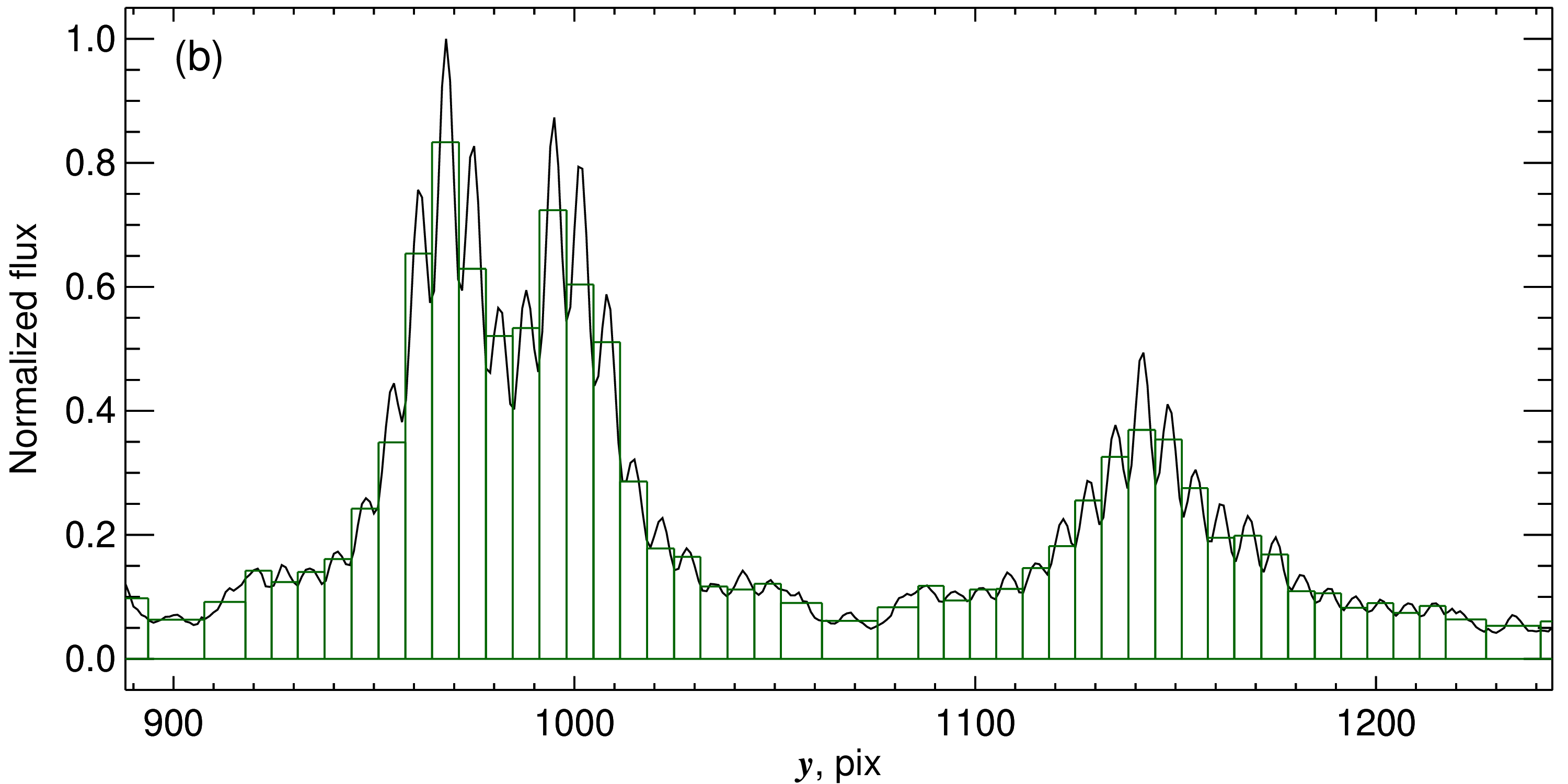}~
	\includegraphics[width=0.495\linewidth]{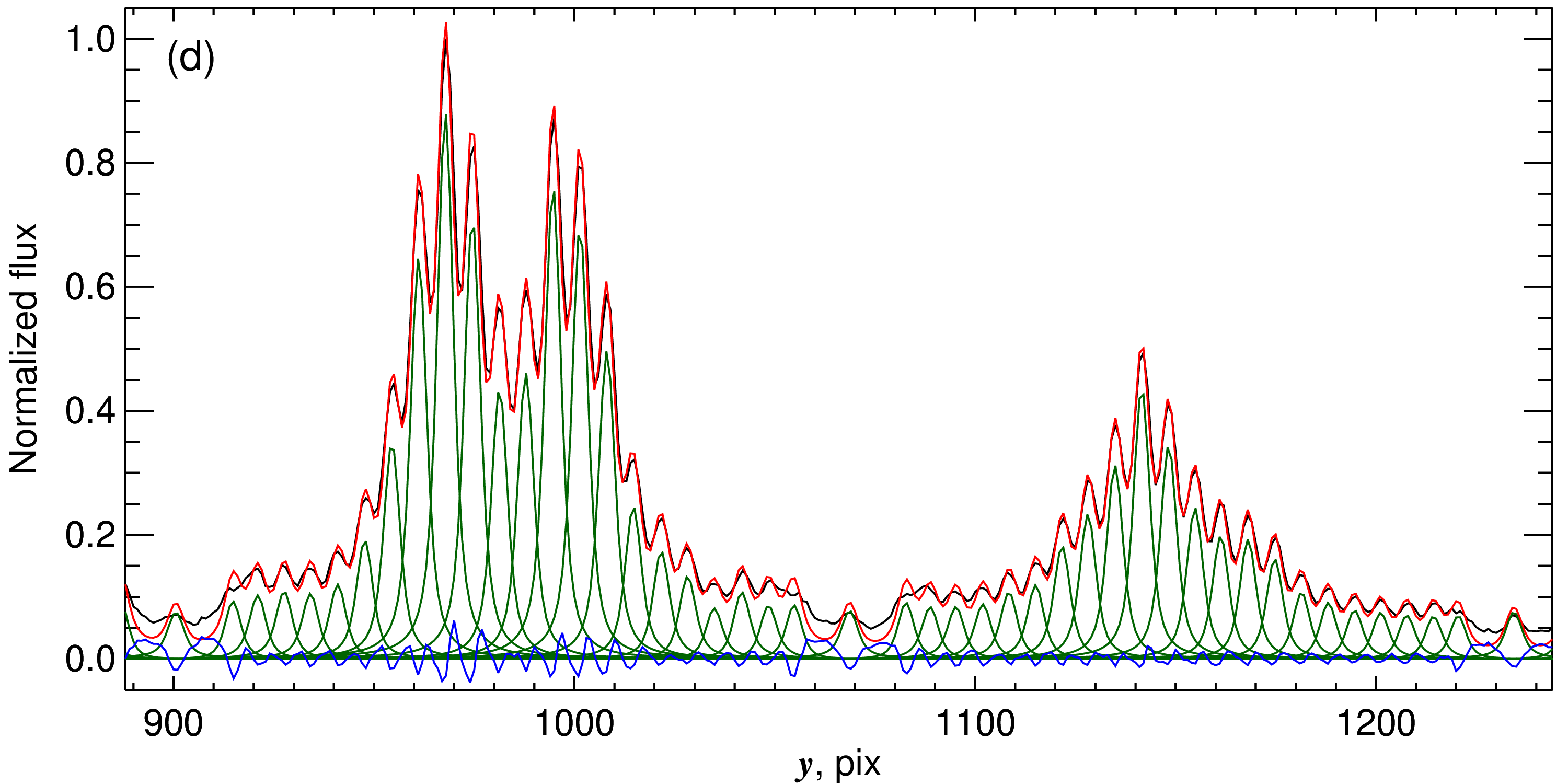}
	\caption{Distribution of normalized intensity across the
		dispersion direction in the \textsc{flat} (a, c) and \textsc{obj}
		(b, d) spectra from the right slit. The widths and areas of
		rectangles in panels (a, b) correspond to the size of the
		integration domain and the flux from each fiber in the case of
		simple extraction of spectra. The green lines in panels (c, d)
		show the intensity distributions from each fiber, the red line
		shows the integrated model spectrum, and the blue line shows the
		difference between the observed and extracted intensity
		distributions for optimum extraction.}
	\label{fig:extract:Afanasiev_n_en}
\end{figure*}

\begin{figure*}
	\includegraphics[width=0.495\linewidth]{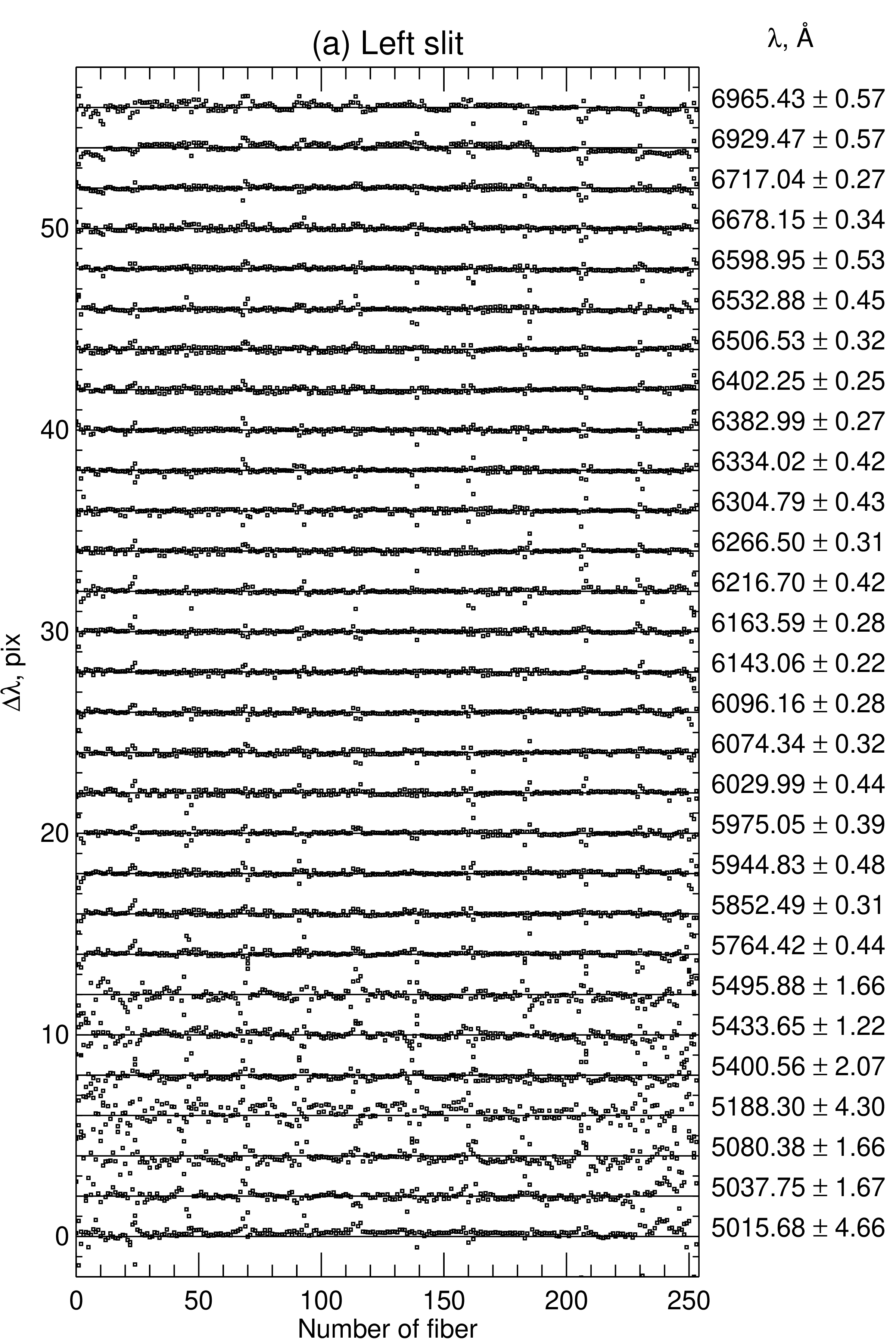}~
	\includegraphics[width=0.495\linewidth]{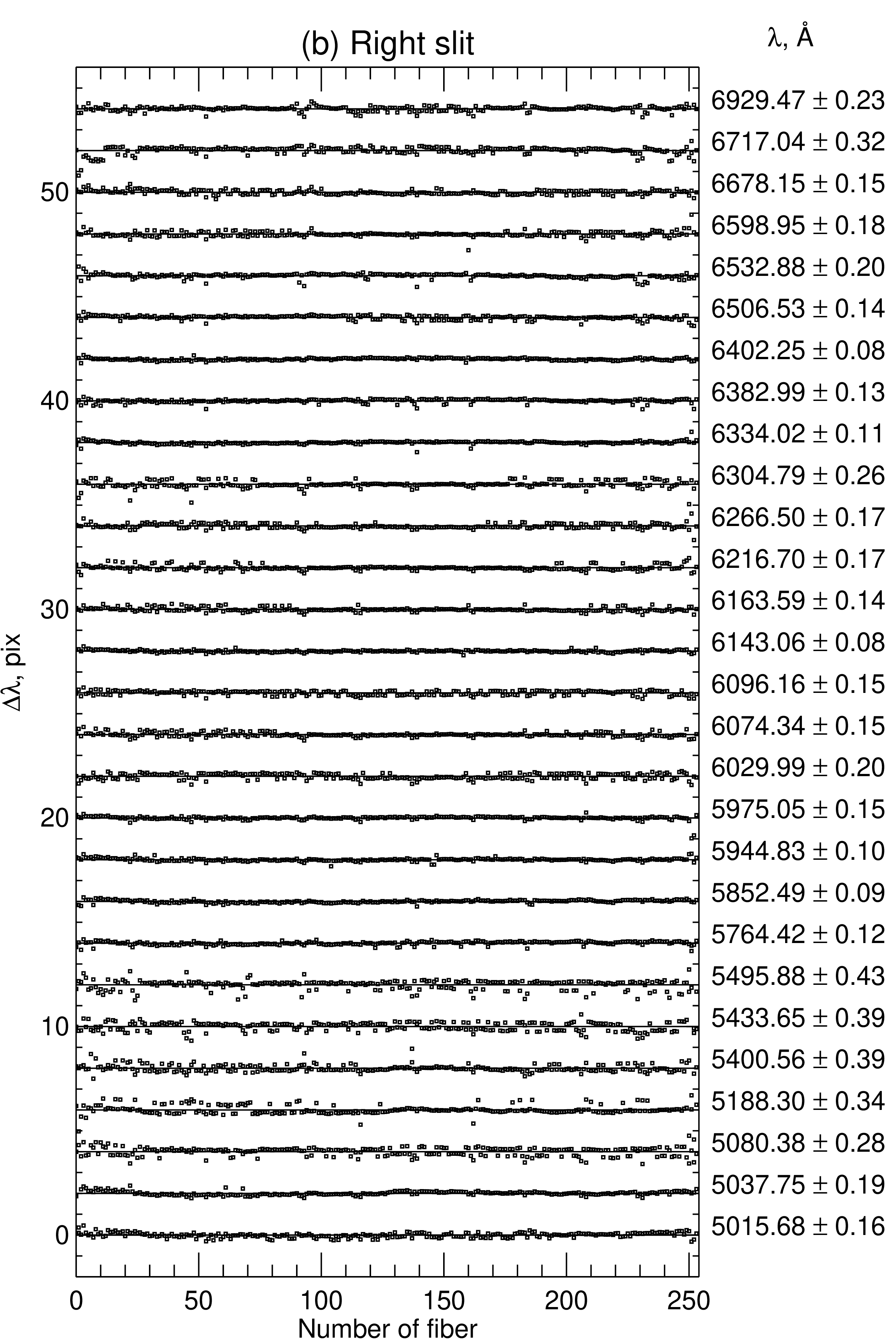}
	\caption{Distribution of deviations of the positions of identified
		lines of the He-Ne-Ar lamp from the laboratory wavelengths after
		linearization procedure along the (a) left and (b) rightslits.}
	\label{fig:lin:Afanasiev_n_en}
\end{figure*}

Search for lines in  \textsc{neon} frames is performed by
identifying peaks in the intensity distribution in the spectrum
integrated along the trajectory of the central  ``sky'' fiber. The
search for peaks can be performed both automatically and via cross
correlation with the model spectrum of the He-Ne-Ar lamp
constructed for each grating used. Then iterative search of line
centers is performed in the spectra obtained by integrating along
the remaining  ``sky'' fibers---the coordinates determined from
the previous spectrum are used as the initial approximation. The
coordinates of each line determined via this procedure are
approximated by a second-order polynomial.

The points of intersection of identified trajectories of ``sky''
fibers (the red lines in Fig.~\ref{fig:geometry:Afanasiev_n_en}b) and the lines
of the He-Ne-Ar lamp (the orange lines in
Fig.~\ref{fig:geometry:Afanasiev_n_en}b) are used as the grid points of the
geometric model of 2D spectra (the magenta points in
Fig.~\ref{fig:geometry:Afanasiev_n_en}b). They are supplemented with a number of
boundary points on each side of the frame obtained by
extrapolating the resulting grid beyond the spectrum. These
operations are followed by the correction of geometric distortions
using  \textsc{warp\_tri}, which is a standard IDL procedure.
Fig.~\ref{fig:geometry:Afanasiev_n_en}c shows the result of correction for
\textsc{neon}.

\subsection{Extraction of Spectra}
\label{sec:red_extract:Afanasiev_n_en}

In the resulting frames the flux from each fiber is not
constrained to a single pixel along the slit but is scattered over
several neighbouring pixels. The aim of the procedure of extraction
of spectra is to obtain the integrated spectrum from each fiber.

To accomplish this stage we use the file containing laboratory
measurements of the positions of all fibers along each slit. Given
the \textsc{eta} frame coordinates of the  ``sky'' fibers
identified at the previous stage we use interpolation to compute
the coordinates of the centers of the trajectories of all fibers
in the \textsc{flat} frame.

Extraction of spectra in IFURED program can be performed in two
ways. Simple extraction consists in integrating the flux within
rectangular aperture around the identified centers of the fiber
trajectories in the 2D spectrum. The halfwidth of the aperture is
set equal to half the distance between the adjacent peaks.
Fig.~\ref{fig:extract:Afanasiev_n_en}a,b illustrates the application of this
technique to the  \textsc{flat} and \textsc{obj} spectra. The
areas of rectangles in this figure corresponds to the integrated
flux in the aperture.

The shortcoming odf simple extraction method is that it does not
account for scattering of light from each fiber: point spread
functions (PSF) of individual fibers overlap.  Part of the
scattered light is a result of scattering in the optics of the
instrument. Unlike MPFS, where scattering amounts to 10--15\%
because of the use of a mirror-lens camera, scattering in
SCORPIO-2 does not exceed 2\%, which is achieved as a result of
application of  AR coatings to optical surfaces and the use of
volume phased holographic gratings. We take into account
scattering via the method of optimum extraction. To this end, we
first construct a model of the flux distribution from fibers along
the slit. The scattering function at the exit of the fiber can be
described quite well by the Voigt function, which has extended and
low-contrast wings. The width of the PSF of each fiber is
determined by fitting the intensity distribution along the slit in
\textsc{flat} frames. To take into account the variation of the
PSF width, this procedure is applied to several dozen sections
along the slit at various wavelengths. The inferred PSF widths of
each fiber are interpolated over the entire spectral range and
fixed for all data types. Final extraction of spectra for all data
types is performed by fitting 254~Voigt profiles with fixed
centers and widths to the intensity distribution along the slit
for each pixel along the dispersion. The only free parameters are
the intensities of all profiles. Fig.~\ref{fig:extract:Afanasiev_n_en}c,d
illustrates the application of optimum extraction procedure to
\textsc{flat} and \textsc{obj} data.

\subsection{Linearization}

The data obtained are wavelength calibrated (linearization of the
spectrum) using the calibration spectrum of the  He-Ne-Ar lamp
(\textsc{neon}). The positions of lines in the  \textsc{neon}
frame are determined using  the cross-correlation method with the
preconstructed model emission spectrum for each employed grating.
The positions of identified lines are used to construct the
dispersion curve---the dependence of the wavelength in the
spectrum on the coordinate in the frame. IFURED allows setting the
degree of the polynomial used to fit the curve both along the
dispersion axis and along the slit to correct for eventual
geometric distortions. The resulting dispersion curve is used to
linearize the entire data set.

Fig.~\ref{fig:lin:Afanasiev_n_en} shows the deviations of the positions of
identified lines of the  He-Ne-Ar lamp from the laboratory
wavelengths after applying the linearization procedure using the
dispersion curve in the form pf a third-order polynomial along the
dispersion and second-order polynomial along the slit. Note that
the large scatter in the blue part of the spectrum is due to low
signal-to-noise ratio in this part of the spectrum, which, in tun,
is due to the short exposure of the  \textsc{neon} frame in the
data set used for this demonstration.

\subsection{Flatfield Normalization and Creation of Data Cubes}
Correction of the observed spectra for nonuniform transmission of
each spectrum due to optical vignetting across the field of view
and variations of the fiber transmission is a necessary stage in
the data reduction. By default IFURED uses to this end the
\textsc{~flat} ``flatfield'' frame --- the spectrum of the (quartz
or LED) flatfield lamp taken with the {SCORPIO-2 calibration
system~\citep{Afanasiev2017:Afanasiev_n_en}}. Nonuniform transmission is corrected
after assembling data cubes.

The data cube is a three-dimensional array with two space
coordinates and one spectral coordinate. They are assembled in
accordance with the given scheme of fiber packing in the IFU unit.
Linearized spectra from the left and right slits form the upper
and lower halves of the $22\times24$~pix$^2$ data cubes. The top
and bottom rows are made of night-sky spectra (NS). These rows are
formed from the data of fewer fibers than pixels and therefore to
ensure uniform their filling the spectra from the NS are written
to every second pixel in the row and the intensity distribution in
other pixels is replaced by averaged spectra of the neighboring
pixels.

The the wavelength-smoothed \textsc{flat} data cube is formed. The
\textsc{flat} spectrum is not flat and therefore \textsc{flat} has
to be normalized to the spectral brightness distribution of the
flat field. We determine it as the brightness distribution
averaged over all spectra in the cube. Every smoothed spectrum in
the \textsc{flat} cube is then normalized to the brightness
distribution of the flat field. All other data cubes are corrected
for nonuniform illumination by dividing them by the normalized 3D
flatfield spectrum described above.

In the cases where we obtain twilight spectra \textsc{sunsky}, we
can use them to validate the quality of the correction of
nonuniformities in the data cubes using normalized
\textsc{flat}---possible errors are due to the violation of the
telecentrism condition in the calibration path. In this case the
corrected  \textsc{sunsky} cube is used to produce the secondary
normalized flatfield (in accordance with the above algorithm) by
which to divide the \textsc{obj} and \textsc{star} data cubes.

\subsection{Subtraction of Night-Sky Lines}
To subtract the night-sky spectrum from all  \textsc{obj} and
\textsc{star} frames the spectra from  ``sky'' fibers are used
that correspond to empty areas aside from the observed object. In
the assembled data cubes they form the upper and lower rows for
the left and right slits, respectively. For each column in the
data cube the spectrum from the  ``sky'' fiber from the same
column is used; the upper and lower rows are used for the upper
and lower half, respectively.

\begin{figure}[t]
    \includegraphics[width=0.95\linewidth]{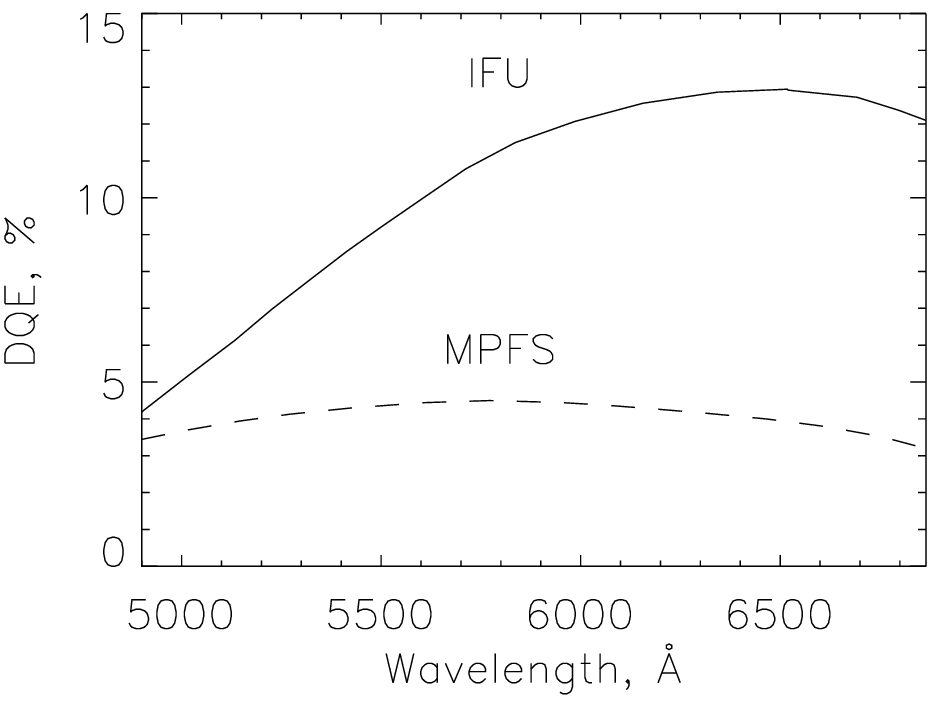}
    \caption{Measured quantum efficiency of  IFU SCORPIO-2 with VPHG940@600 grating. For comparison the dashed line
    shows the quantum efficiency of MPFS spectrograph with 600~mm$^{-1}$ reflective diffraction grating.}
    \label{fig:dqe:Afanasiev_n_en}
\end{figure}

\subsection{Flux Calibration, Correction of the Atmospheric Extinction and Dispersion}
\begin{figure*}
    \includegraphics[width=1\linewidth]{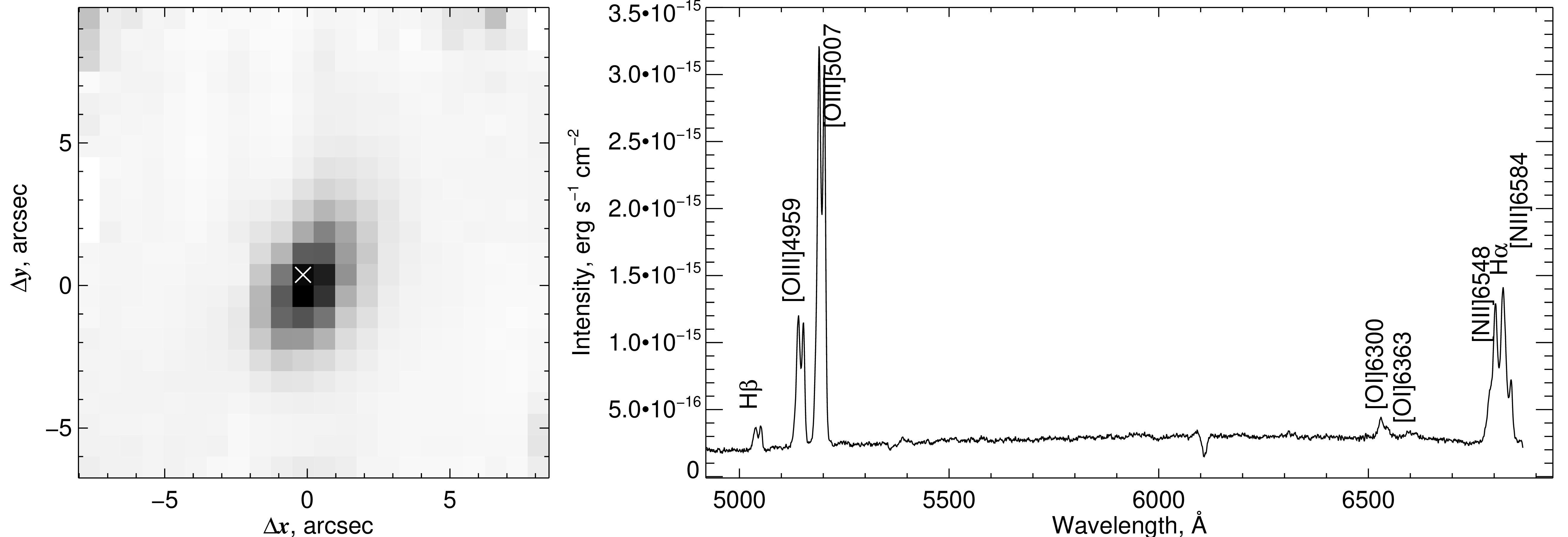}
    \caption{Integrated image of Mrk\,78 in the continuum. The right panel shows, by way of an example, the spectrum corresponding
    to the pixel marked by the red cross, and also indicates the identification of observed bright emission lines.}
    \label{fig:cub:Afanasiev_n_en}
\end{figure*}
Calibration of fluxes in terms of energy units is performed in the
standard way---the observed integrated spectrum \textsc{star} of
one of the spectrophotometric standards (e.g., adopted from the
list published in~\citealt{Oke1990:Afanasiev_n_en}) is compared to
the known spectral energy distribution for this object. Given the
measured coefficients of atmospheric extinction for the Special
Astrophysical Observatory of the Russian Academy of
Sciences~\citep{extin2:Afanasiev_n_en,extin1:Afanasiev_n_en}, we
can compute the extinction at the zenith angle corresponding to
the observed standard. After that it is easy to directly estimate
the expected number of photons from the spectrophotometric
standard  incident onto the telescope mirror and compare the
result with the observed values at different wavelengths. The
resulting quantum efficiency curve (DQE) (see
Fig.~\ref{fig:dqe:Afanasiev_n_en}) is used to take into account
the spectral response of the instrument. The fluxes in the
observed \textsc{obj} data cube are then multiplied by the
transformation coefficient from counts to energy units computed
from the \textsc{star} cube and also by the normalization
coefficient taking into account  the difference the zenith angles
of the object and spectrophotometric standard and the times of
their exposures. As is evident from the figure, DQE\,$\sim$\,13\%
at 6000~\AA\ in the IFU mode with VPHG940@600 grating. For the
same grating our measurements in the long-slit mode yield
DQE\,$\sim$\,42\%~\citep{Sco2:Afanasiev_n_en}. The result of a
comparison of the LS and IFU modes shows that in the latter case
the quantum efficiency of the instrument decreases by a factor of
about 3. This conclusion is consistent with our estimates
mentioned in Section~2.2, where we conclude that the quantum
efficiency decreased by a factor of~2.5. For comparison, the
dashed line in the same figure shows the DQE curve for  MPFS
spectrograph obtained with a 600~lines~mm$^{-1}$ grid. Note that
we used the data obtained with a similar dispersion:
1.2~\AA~mm$^{-1}$ for IFU and 1.42~\AA~mm$^{-1}$ for MPFS. The
maximum DQE for MPFS is of about~4.5\%, whereas this is almost
three times less than our achieved efficiency in the  IFU mode.
The efficiency of the IFU mode when operating with a higher
dispersion (VPHG with a modulation of 1800~and 2300mm$^{-1}$)
decreases to about~6\%.

The last stage of data reduction is the correction for atmospheric
dispersion. To this end, the wavelength dependence of the position
of the barycenter of the  \textsc{obj} object in the field of view
is constructed, which is used to correct the \textsc{obj} file. At
zenith angles smaller than~50\degr the corresponding offsets do
not exceed one pixel throughout the entire spectral range.

\begin{figure*}
    \includegraphics[width=1\linewidth]{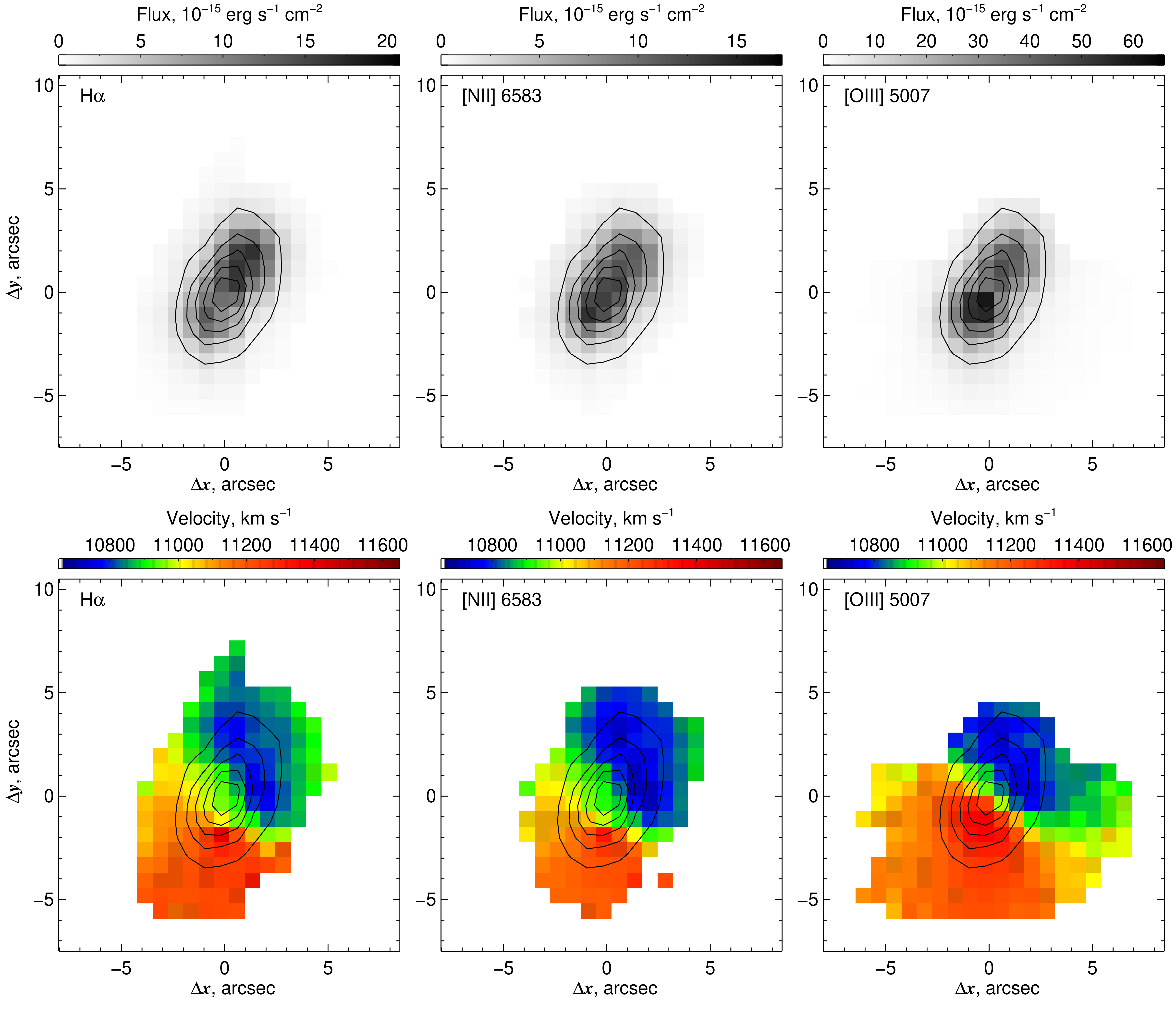}
\caption{Result of fitting   H$\alpha$, [NII]\,6584~\AA\, and
[OIII]\,5007~\AA\ lines (left to right) by a single-component
Gaussian profile and continuum. Top to bottom: distribution of
line flux and radial-velocity fields. The contours show the
brightness distribution in the continuum.}
    \label{fig:maps:Afanasiev_n_en}
\end{figure*}

\subsection{Analysis of Reduced Data}

In addition to the reduction procedures described above  IFURED
also includes tools for data vizualization and basic analysis.
Fig.~\ref{fig:cub:Afanasiev_n_en} shows the image of  Mrk\,78 galaxy in the
continuum obtained from the assembled data cube, and also an
example of the spectrum corresponding to a pixel located near the
center of the galaxy. As is evident from the figure, the use of
VPHG940@600 grating with IFU allows studying emission lines in the
wavelength interval from H$\beta$ to [NII]\,6584~\AA\,(up to
[SII]\,6731~\AA in the case of closer objects).

To illustrate the use of IFURED or basic analysis of the data,
Fig.~\ref{fig:maps:Afanasiev_n_en} shows the flux and radial-velocity maps of
Mrk\,78 galaxy in H$\alpha$, [NII]\,6584~\AA\, [OIII]\,5007~\AA\,
and H$\beta$ lines. These images were acquired bt approximating
each emission line by a single-component Gaussian profile and
underlying second-order polynomial describing the continuum via
\textsc{mpfit}~\citep{mpfit:Afanasiev_n_en} procedure. The  [OIII]\,4959,5007\AA\, and
[NII]\,6548,6584\AA\, line doublets can in this case be approximated
by two components with a fixed separation and the same FWHM.
The contours in the figure show the continuum brightness
distribution.

In the maps presented here two bright regions in Mrk\,78 can be
seen in all emission lines, whereas in the continuum the galaxy
mostly shows up between these regions. The $H\alpha$ velocity
field shows regular rotation, and its distortion in the [OIII]
line is indicative of noncircular motions.

Note, however, that in the case of Mrk\,78 a single-component
Gaussian profile fit gives only a rough idea about the kinematics
and morphology of gas in the galaxy. Previous observations of
Mrk\,78 revealed complex kinematics of ionized gas, which
manifests itself in observed multicomponent emission-line
profiles~\citep{Whittle2004:Afanasiev_n_en}. Bifurcation of line
profiles can also be clearly seen in our
Fig.~\ref{fig:cub:Afanasiev_n_en}. A detailed analysis of the
kinematics and gas ionization mechanisms in  Mrk\,78 galaxy
requires a special paper.

\section{CONCLUSION}
The multimode  SCORPIO-2 spectrograph designed for the 6-m
telescope includes an IFU unit with a lens raster meant for field
spectroscopy. Here are the main points:
\newcounter{N}
\begin{list}{\arabic{N}.}{
\setlength\leftmargin{5mm} \setlength\topsep{2mm}
\setlength\parsep{0mm} \setlength\itemsep{2mm} }
\usecounter{N}
\item The operation principle of IFU---a square raster made of
microlenses with  $23^\times $ magnifying optics. \item The lens
raster contains   $22\times22$ square microlenses, each with the
size of 2~mm; the image scale is $0\farcs75$/lens, and the size of
the field of view is  $16\farcs5\times16\farcs5$. \item Optical
fibers reform micropupil images into two pseudoslits located at
the IFU collimator entrance. Each slit contains 254 fibers  (242
fibers from the object + 12 fibers from the sky background). At
the spectrograph output two arrays is formed, each consisting of
254 spectra. \item A set of volume phased holographic gratings
(VPHG) provides the spectral interval 4600--7300~\AA\ and a
resolution of $\lambda/\delta\lambda$ from 1040 to 2800 in the IFU
mode. \item The quantum efficiency of the  IFU mode is 6--13\%
depending on the grating employed, which, all other conditions
being equal, exceeds the efficiency of MPFS by a factor of about
three.
 \item We developed IFURED software package for primary
reduction of the data.
\end{list}

The IFU mode in SCORPIO-2 spectrograph is meant for field
spectroscopy of central regions in galaxies in the H$\beta$ and
H$\alpha$ lines, for diagnosing the conditions of gas ionization
and studying of its kinematics. We also plan to study the stellar
component in central parts of galaxies in absorption lines---the
chemical composition, stellar velocity dispersion and stellar
velocity field. We consider this to be an extra mode in addition
to photometry, long-slit spectroscopy, and field spectroscopy with
a Fabry--Perot interferometer for observing extended objects with
SCORPIO-2.

\begin{acknowledgments}
We are grateful to R.~I.~Uklein for measuring flexures on the 6-m
telescope of the Special Astrophysical Observatory of the Russian
Academy of Sciences and to the administration of the Special
Astrophysical Observatory of the Russian Academy of Sciences for
allocating observational time for technical tests. This work was
supported by the Russian Science Foundation (project
no.~17-12-01335).
\end{acknowledgments}



\begin{flushright}
\textit{Translated by A.~Dambis}
\end{flushright}


\end{document}